\documentclass[fleqn]{aa}
\usepackage{mathtools}
\usepackage{txfonts}
\usepackage{natbib}
\usepackage{graphicx}
\usepackage{epstopdf}
\usepackage{pgf}

\usepackage{soul}
\definecolor{lightgreen}{HTML}{B7F774}
\sethlcolor{yellow}

\providecommand{\mc}[1]{\multicolumn{1}{c}{#1}}

\def\fdg{\mbox{$.\!\!^\circ$}}

\begin{document}

\title{{\em RadioAstron} space VLBI imaging of polarized radio emission in the high-redshift 
quasar 0642$+$449 at 1.6\,GHz}

\author{A.~P.~Lobanov\inst{1,2}
        \and
        J.~L.~G\'omez\inst{3}
        \and
        G.~Bruni\inst{1}
        \and
        Y.~Y.~Kovalev\inst{4,1}
        \and
        J.~Anderson\inst{1,5}
        \and
        U.~Bach\inst{1}
        \and
        A.~Kraus\inst{1}
        \and
        J.~A.~Zensus\inst{1}
        \and
        M.~M.~Lisakov\inst{4}
        \and
        K.~V.~Sokolovsky\inst{4,6}
        \and
        P.~A.~Voytsik\inst{4}
}

    \institute{Max-Planck-Institut f\"ur Radioastronomie,
              Auf dem H\"ugel 69, 53121 Bonn, Germany
    \and
    Institut f\"ur Experimentalphysik, Universit\"at Hamburg, 
    Luruper Chaussee 149, 22761 Hamburg, Germany
    \and
    Instituto de Astrof\'isica de Andaluc\'ia - CSIC, 
    Glorieta de la Astronom\'ia s/n, E-18008 Granada, Spain
    \and
    Astro Space Center of Lebedev Physical Institute, Profsoyuznaya 84/32, 117997 Moscow, Russia
    \and
    Helmholtz-Zentrum Potsdam, Deutsches GeoForschungsZentrum GFZ, Telegrafenberg A6, 14473 Potsdam, Germany
    \and
    Sternberg Astronomical Institute, Moscow State University, Universitetskii~pr. 13, 119992 Moscow, Russia
  }

\authorrunning{Lobanov et al.}
\titlerunning{RadioAstron polarization image of 0642$+$449}

  \date{}
 
  \abstract
  % context heading (optional), leave it empty if necessary  
  {Polarization of radio emission in extragalactic jets at a
    sub-milliarcsecond angular resolution holds important clues for
    understanding the structure of the magnetic field in the inner
    regions of the jets and in close vicinity of the
    supermassive black holes in the centers of active galaxies.}
  % aims heading (mandatory)
  {Space VLBI observations provide a unique tool for polarimetric
    imaging at a sub-milliarcsecond angular resolution and studying the
    properties of magnetic field in active galactic nuclei on scales
    of less than $10^4$ gravitational radii.}
  % methods heading (mandatory)
  {A space VLBI observation of high-redshift quasar TXS 0642$+$449 (OH\,471),
    made at a wavelength of 18\,cm (frequency of 1.6\,GHz) as part of
    the Early Science Programme (ESP) of the {\em
      RadioAstron} mission, is used here to test the polarimetric performance
    of the orbiting Space Radio Telescope (SRT) employed by the
    mission, to establish a methodology for making full Stokes
    polarimetry with space VLBI at 1.6\,GHz, and to study the
    polarized emission in the target object on sub-milliarcsecond
    scales.}
  % results heading (mandatory)
  {Polarization leakage of the SRT at 18\,cm is found to be within 9\%
    in amplitude, demonstrating the feasibility of high fidelity
    polarization imaging with {\em RadioAstron} at this wavelength.
    A polarimetric image of 0642$+$449 with a resolution of 0.8\,mas
    (signifying an $\sim 4$ times improvement over ground VLBI
    observations at the same wavelength) is obtained. The image shows
    a compact core-jet structure with low ($\approx 2\%$) polarization
    and predominantly transverse magnetic field in the nuclear region.
    The VLBI data also uncover a complex structure of the nuclear
    region, with two prominent features possibly corresponding to the
    jet base and a strong recollimation shock. The maximum brightness
    temperature at the jet base can be as high as $4\times
    10^{13}$\,K.}
  % conclusions heading (optional), leave it empty if necessary 
   {}


   \maketitle
%
%________________________________________________________________

\section{Introduction}

Very long baseline interferometry (VLBI) observations in which one of
the antennas is placed on board of an Earth satellite (space VLBI) have
provided a capability of reaching unprecedentedly high angular
resolution of astronomical observations
\citep{arsentev+1982,levy+1986,hirabayashi+2000}. The space VLBI
mission {\em RadioAstron} combines ground-based radio antennas
operating at frequencies 0.32, 1.6, 5, and 22 Gigahertz (GHz) with a
10-meter antenna (Space Radio Telescope, SRT) on board of the 
satellite {\em Spektr-R} launched on 18 July 2011 \citep{kardashev+2013}. 

At 0.32, 1.6, and
22\,GHz, the SRT delivers data in both left (LCP) and right (RCP)
circular polarization at 0.32, 1.6, and 22\,GHz, which enables fully
reconstructing the linearly polarized emission from target objects at
these frequencies.  At 5\,GHz, the SRT provides only LCP data because of 
a failure of the on-board hardware.

After its launch, the SRT was first tested in a single-antenna mode during
the in-orbit-checkout (IOC) period \citep{kovalev+2014} and later in
the interferometry mode, in combination with ground-based antenna
(fringe search, FS), successfully delivering fringes at each of the
four observing bands \citep{kardashev+2013,kardashev+2014}. 

The ensuing early science program (ESP) of {\em RadioAstron} started
in February 2012 and continued through June 2013
\citep{kardashev+2014}.  One of the prime objectives of the program
was to provide a bridge between the initial experimental mode of
operations, observing, and data processing, and routine operations
that started after completion of the ESP.  

One of the main goals of the ESP was to establish the feasibility of
imaging experiments with {\em RadioAstron}. The first imaging
experiment with {\em RadioAstron}, targeting the bright and compact
radio source 0716+714, was made during the ESP period at 5\,GHz
\citep{kardashev+2014}. Polarization imaging was tested with the
observation presented in this paper.

Polarization imaging at high angular resolution, both along and
transverse to the jet direction, is crucial for measuring the magnetic
field structures of jets to distinguish between different possible
magnetic field configurations. The improvement of angular resolution
provided by space VLBI observations makes it particularly interesting
to compare the brightness temperature and polarization measurements
made with {\em RadioAstron} to the ground VLBI with a similar
resolution obtained at higher observing frequencies.
It has been shown that a minimum of several, independent, resolution
elements across a jet are necessary to prevent spurious results from
being obtained \citep{hovatta+2012}. Space VLBI observations are
essential for resolving transversely a significant number of jets and
studying the structure and properties of their magnetic field.

Feasibility of polarization measurements with space VLBI has been
demonstrated with the {\em VSOP} \citep{hirabayashi+2000}
observations made with a satellite antenna recording data only in
single (left circular) polarization \citep{kemball+2000,gabuzda+2001}.
Results from the first polarization observation with {\em RadioAstron}
at a frequency of 1.6\,GHz are reported here. This observation was
made in March 2013 as part of the ESP and it targeted a bright,
compact radio source TXS~0642$+$449 (OH\,471).

The radio source 0642$+$449 is a distant
low-polarization quasar (LPQ) located at a redshift of 3.396
\citep{osmer+1994} which corresponds to the luminosity distance of
30.2 Gigaparsecs (Gpc) and a linear scale of 7.58\,kpc/arcsecond,
assuming the standard $\Lambda$CDM cosmology \citep{planck2015}. The
optical spectrum of 0642$+$449 does not show strong broad lines
\citep{torrealba+2012}. In radio, its radio structure has a compact
``core--jet'' morphology \citep{gurvits+1992,xu+1995}, with a flat
spectral index $\alpha \approx -0.1$ \citep{hovatta+2014} of the
milliarcsecond (mas) scale core (bright narrow end of the jet).

The maximum brightness temperature, $T_\mathrm{b}$, measured in
0642$+$449 decreases with observing frequency. Previous {\em VSOP}
space VLBI observations of 0642$+$449 at 5\,GHz put a lower limit of
$T_\mathrm{b}\ge 3.3\times 10^{12}$\,K \citep{dodson+2008}, while
ground VLBI observations at 15 and 86\,GHz resolved the
  compact core and yielded somewhat lower brightness temperatures of
1.2--4.3$\,\times 10^{12}$\,K \citep{kovalev+2005} and
1.1--1.6$\,\times 10^{11}$\,K \citep{lobanov+2000,lee+2008}. This
implies that the compact core becomes gradually resolved out at higher
frequencies, calling for comparison with the new space VLBI
observations by {\em RadioAstron}.

Ground VLBI monitoring of the source has revealed a range of proper
motions of 0.012--0.078\,mas/yr in the jet \citep{lister+2013} and
provided estimates of the jet Lorentz factor, $\Gamma_\mathrm{j} =
5.4$, viewing angle $\theta_\mathrm{j} = 0.8^{\circ}$ and intrinsic
opening angle $\phi_\mathrm{j} = 0.3^{\circ}$
\citep{pushkarev+2009}. With these parameters, an equipartition
magnetic field of $\approx 0.3$\,G can be estimated at the location of
the jet core \citep{lobanov1998} observed at 1.6 GHz, based on
measurements of frequency dependent shift of the core position which
have been reported for 0642+449 between frequencies of 8.1, 8.4, 12.1,
and 15.4\,GHz \citep{pushkarev+2012}.

Ground VLBI polarization observations at 5.0 and 8.4\,GHz
\citep{osullivan+2011} have shown a small (1.1--2.6\%) degree of
polarization in the core, with no conclusive evidence for Faraday
rotation ($\chi = 15.6^{\circ}$ and $15.3^{\circ}$ at 5.0 and
8.4\,GHz, respectively). At higher frequencies, between 8.4 and
15.1\,GHz, a rotation measure (RM) of $-280$ rad/m$^2$ has been
reported in the core \citep{hovatta+2012}. The Galactic RM in the
direction of 0642$+$449 is estimated to be $-5.4 \pm 18.3$ rad/m$^2$ \citep{taylor+2009,oppermann+2012}.

The polarization observations of 0642$+$449 with {\em RadioAstron} signal
the first successful space VLBI polarization experiment using dual
circular polarization measurements made at the space-borne
antenna. The methodology for {\em RadioAstron} polarimetric observations and
data correlation is presented in Sect.~\ref{sc:method}. The
{\em RadioAstron} observation of 0642$+$449 and reduction of the
polarimetric data are described in Sect.~\ref{sc:reduction}.  The
polarization image of 0642$+$449 is discussed in Sect.~\ref{sc:image}
and compared with previous ground VLBI observations of this object.

\section{Imaging with {\em RadioAstron}}
\label{sc:method}

{\em RadioAstron} observations are performed by combining together
ground-based radio telescopes with the orbiting 10-meter antenna of
the SRT. The data are recorded at a rate of 128\,Megabits per second,
providing a total bandwidth of up to 32\,MHz per circular polarization
channel \citep{andreyanov+2014}. The satellite telemetry and science
data are received and recorded at tracking stations in Puschino
\citep{andreyanov+2014} and Green Bank \citep{ford+2014} and later
transferred for correlation at one of the {\em RadioAstron}
correlation facilities. Accurate reconstruction of the orbit and of
the momentary state vector of the {\em Spektr-R} spacecraft is
achieved by combining together radiometric range and radial
  velocity measurements made at the satellite control stations in Bear
  Lakes and Ussurijsk (Russia), Doppler measurements obtained at each
  of the tracking stations, and laser ranging and optical astrometric
  measurements of the spacecraft position in the sky
\citep{khartov+2014,zakhvatkin+2014}. The accuracy can be further
enhanced by simultaneous VLBI measurements performed on the
narrow-band downlink signal from the spacecraft
\citep{duev+2012}. This was demonstrated \citep{duev+2015} using the
data from the ESP observations of 0642$+$449 discussed here.

The {\em Spektr-R} satellite has a highly elliptical orbit, with the
apogee reaching up to 350,000\,km and the perigee varying in the
7,000--80,000\,km range. The orbital period of the satellite varies
between 8.3 and 9.0 days \citep{kardashev+2013,kardashev+2014b}.
These factors make imaging observations with {\em RadioAstron}
particularly challenging, with optimal conditions realized typically
only for a limited fraction of sky and for limited periods of time.
The imaging observations are planned by simulating the observing
conditions in the {\em FakeRaT} software
package\footnote{http://www.asc.rssi.ru/radioastron/software/soft.html}
\citep{zhuravlev2014}, which is based on the {\em FakeSat} software
\citep{murphy1991,murphy+1994}.

The Fourier spacings provided by the ground-space baselines to the SRT
are typically limited to a very narrow range of position angle and
often does not reach down to the spacings of the baselines between
ground-based telescopes. In order to reduce the adverse effect of
these two factors, imaging with {\em RadioAstron} is performed during the
perigee passages of the SRT ({\em perigee imaging}), typically with
12-24 hour-long observing segments. Whenever feasible, these perigee
imaging segments are augmented by several 1-2 hour-long segments ({\em
  long-baseline tracking}) allocated during the following days and
including a smaller number (3-4) of ground antennas. The long-baseline
tracking provides information on largest Fourier spacings, enabling
detecting the emission on the finest angular scales.

\subsection{Correlation of {\em RadioAstron} experiments}
\label{sc:difx}

Correlation of {\em RadioAstron} experiments is performed at
three different correlator facilities: the ASC {\em RadioAstron} correlator
\citep{kardashev+2013,andrianov+2014}, the JIVE SFXC correlator \citep{kettenis2010},
and the DiFX software correlator \citep{deller+2007,deller+2011}.  In
order to enable the correlation of {\em RadioAstron} data to be made in the
DiFX correlator, the DiFX code has been upgraded by the
Max-Planck-Institute for Radio Astronomy (MPIfR) in Bonn, addressing
the specific aspects of data and telemetry formats of the SRT telescope on-board the {\em Spektr-R} spacecraft
\citep{bruni2014,bruni+2015}.

The upgrade has comprised a new routine to read the native {\em
  RadioAstron} Data Format (RDF) into the DiFX correlator and a number
of modifications to the correlator code enabling it to process data
recorded at the SRT.  The delay model server CALC from the CALC/SOLVE
package\footnote{http://gemini.gsfc.nasa.gov/solve/} used by the DiFX
correlator is modified to be able to calculate delay information for
telescopes with arbitrary coordinates and velocities, accounting for
the general relativistic effects in the gravitational field of the
Earth. The DiFX metadata system is upgraded to
deal with the changing position and velocity of the spacecraft as a
function of time.

Satellite orientation parameters provided in the telemetry are used in
the upgraded version of the code for calculating the equivalent of
parallactic angle correction for the spaceborne antenna. The estimated
accuracy of the orientation parameters is $\approx 1^{\prime\prime}$
\citep{lisakov+2014}. This enables accurate calibration of the
instrumental polarization of the SRT.

The timestamps for the SRT data (set by the tracking station clock at
the moment it begins recording individual VLBI scans) are modified to
refer to the relative time of arrival of the astronomical signal at
the spacecraft. This is done by calculating the delay for the
transmission of the signal from the spacecraft to the tracking
station.  These corrections are inserted to the metadata handler and
to the delay model server, which provides the full compatibility of
the {\em RadioAstron} data stream with data streams from ground antennas.

The upgraded DiFX correlator is now used at the MPIfR to correlate
{\em RadioAstron} data and being prepared for merging with the {\em
  trunk} version of the DiFX correlator code.

% Figure 1: uv-plot
\begin{figure}[ht!]
  \centering
\includegraphics[width=0.45\textwidth]{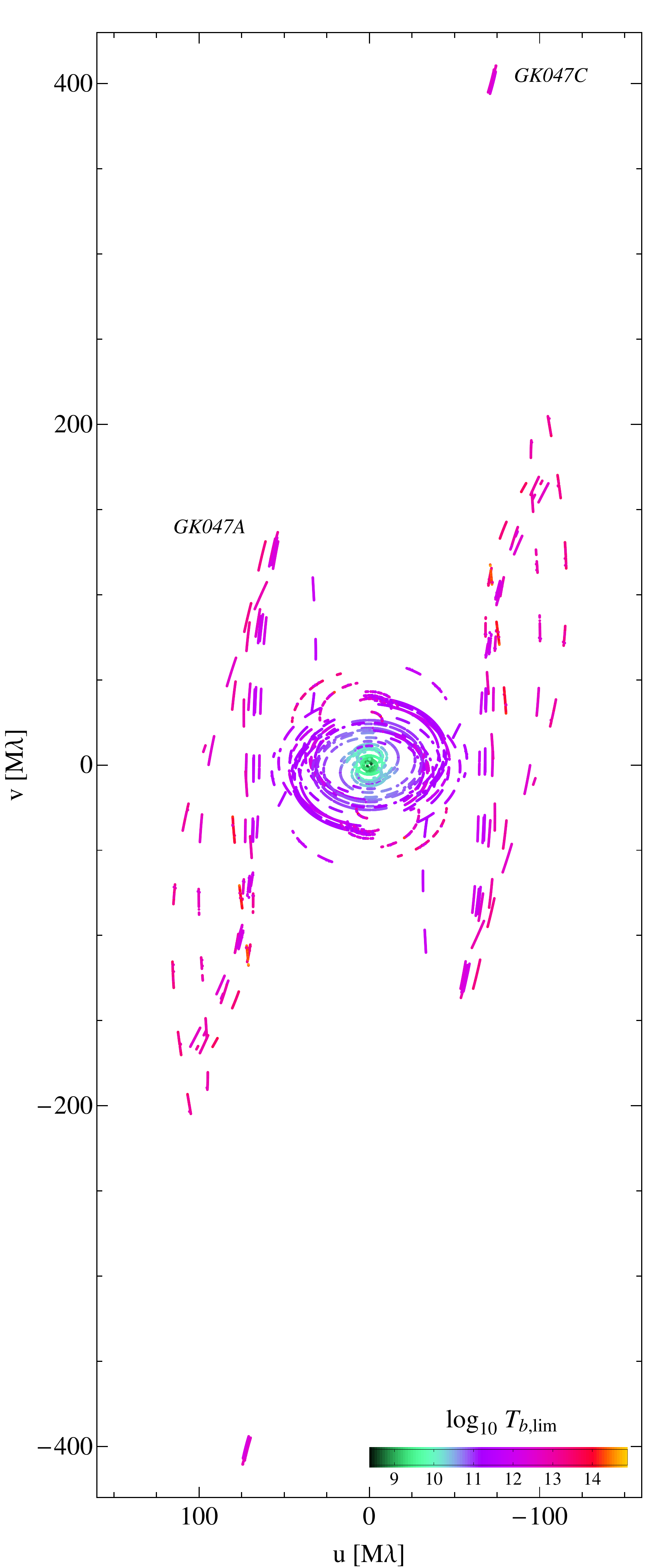}
\caption{Coverage of the Fourier domain ({\em uv} coverage) of the
  {\em RadioAstron} observation of 0642$+$449 at 1.6\,GHz ($\lambda =
  18$\,cm), plotted in units of M$\lambda$. The central part
  corresponds to the perigee imaging segment (GK047A) and the short
  segment at $v \approx 400$\,M$\lambda$ represents the long-baseline
  tracking segment (GK047C). Color marks the
  upper limit of brightness temperature obtained from visibility amplitudes
  (for details, see discussion in Sect.~\ref{sc:tb}).}
\label{fg:0642-uvplot}
\end{figure}
%t l b r: angle=-90
%l b r t: angle=0

\begin{table*}[ht!]
\caption{Radio telescopes participating in {\em RadioAstron} observation of 0642$+$449}
\label{tb:telescopes}
\begin{center}
\begin{tabular}{rc|ccccc}\hline\hline
Telescope      &    & $D$ & SEFD   & \multicolumn{2}{c}{Observing time [UT]} & $B_
\mathrm{max}$ \\
               & Code   & [m] & [Jy]   & GK047A       &  GK047C      & [$D_\mathrm{E}$]  \\\hline
{\em Spektr-R} SRT (RU)   & RA & 10 & 2840 & 10:00 -- 01:00 & 14:00 -- 15:00  & ...  \\
Effelsberg (DE)     & EF & 100& 19 & 10:00 -- 01:00 & 14:00 -- 15:00& 5.87 \\
Jodrell Bank (UK)   & JB & 76 & 65 & 10:00 -- 01:00 & 14:50 -- 15:00& 5.93 \\
WSRT (NL)           & WB & 66$^{\dag}$ & 40 & 13:00 -- 01:00 & 14:00 -- 15:00& 5.90 \\
Torun (PL)          & TR & 32 & 300& 10:00 -- 01:00 &               & 2.76 \\
Urumqi (CH)         & UR & 25 & 300& 10:00 -- 23:00 &               & 3.02 \\
Shanghai (CH)       & SH & 65 & 670& 10:00 -- 19:00 &               & 3.26 \\
Noto (I)            & NT & 32 & 784& 10:30 -- 01:00 &               & 2.69 \\
Hartebeesthoek (SA) & HH & 26 & 430& 14:30 -- 20:00 &               & 1.91 \\
Green Bank (USA)    & GB & 100$^\dag$& 10 & 17:00 -- 01:00 &               & 1.63 \\
Zelenchukskaya (RU) & ZC & 32 & 300& 19:30 -- 23:30 &               & 1.67 \\\hline
\end{tabular}
\end{center} {\bf Notes:} $D$ -- antenna diameter ($\dag$ --
equivalent diameter); SEFD -- system equivalent flux density (system
noise), describing the antenna sensitivity; $B_\mathrm{max}$ --
largest projected baseline in the data, in units of the Earth
diameter.
\end{table*}

\section{{\em RadioAstron} observation of 0642$+$449}
\label{sc:reduction}

The 1.6\,GHz {\em RadioAstron} observation of 0642$+$449 (RadioAstron
project code {\em raes12}; global VLBI project code GK047) was made on
9-10 March 2013, with three observing segments allocated for the
experiment.  The main perigee imaging segment, GK047A (raes12a;
10:00\,UT/09/03--01:00\,UT/10/03) was complemented by VLBI
observations of the satellite downlink signal at 8.4\,GHz (segment
GK047B) made for the purpose of improving the {\em Spektr-R} orbit
determination \citep{duev+2015}. The third segment, GK047C (raes12b;
14:00\,UT/10/03--15:00\,UT/10/03) was introduced as a long-baseline
tracking segment. The SRT data and the satellite telemetry were
recorded by the {\em RadioAstron} satellite tracking station in
Puschino.  Basic parameters of the ground telescopes participating in
the segments GK047A and GK047C are given in Table~\ref{tb:telescopes}.

% Figure 2: SRT residual rate
\begin{figure}[h!]
%\centerline{\includegraphics[width=0.48\textwidth]{resdel-nsec.pdf}} 
\centerline{\includegraphics[width=0.485\textwidth]{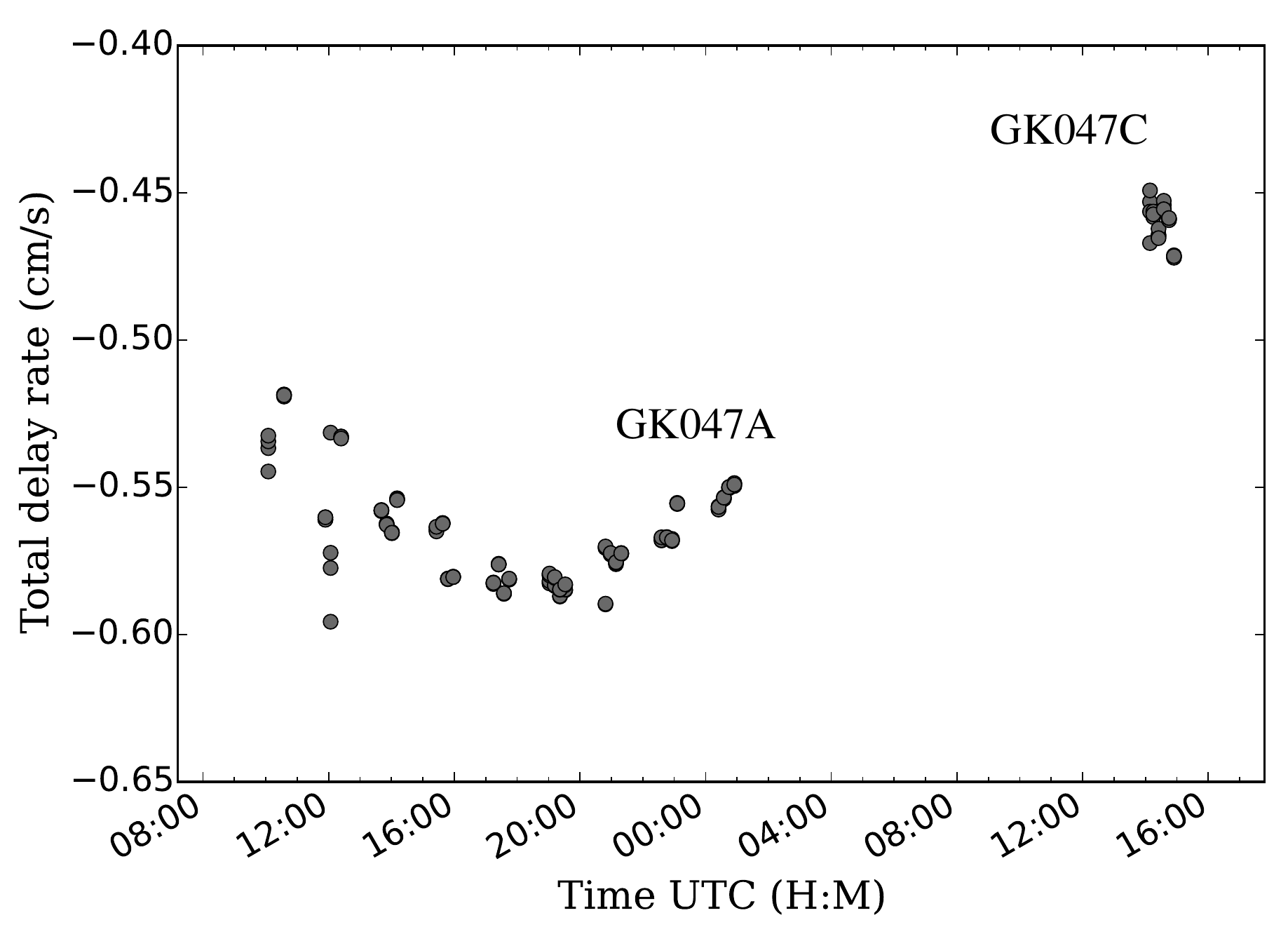}}
\caption{Total delay rates of the SRT obtained from baseline-based
  fringe fitting of the correlated data on the baseline to Effelsberg
  and accounting for the clock rates at each of the two
  telescopes. For each SRT observing segment, the rates are obtained
  from fringe fitting solutions with a time interval of 10 minutes. The
  rates are plotted separately for each IF and polarization
  channels. The plot reflects the accuracy of the velocity
  determination in the reconstructed orbit of the SRT.  All of the
  velocity and acceleration values correspond to the baseline
  projections of the true residual offset, velocity, and acceleration
  of the spacecraft.}
\label{fg:fringe}
\end{figure}

%  The largest
%  delay rates of $\approx 2.5$\,cm/s are obtained at the beginning of
%  the experiment GK047A, shortly after the perigee passage. Dashed
%  lines represent piecewise (UT 10--21, UT 21--14) fits to the delay
%  rates. The polynomial fit implies an initial residual deceleration
%  of $\approx 2\,\mu$m/s$^2$. 

The data were recorded in two polarization channels (left and right
circularly polarized, LCP and RCP), with a bandwidth of 32 MHz per
channel, centered at a frequency of 1.6601875\,GHz. In each of
the channels, the total bandwidth was split into two independent
intermediate frequency (IF) bands of 16 MHz.

The space VLBI segments on 0642$+$449 were observed in a 40/120 minute on/off
cycle, with the gaps between the segments required for cooling the motor
drive of the on-board high-gain antenna of the {\em Spektr-R}.  During these
gaps, the strong radio sources 1328+207 (3C~286) and 0851+202 (OJ~287)
were also observed with the ground antennas as polarization
calibrators.  Fig.~\ref{fg:0642-uvplot} shows the resulting visibility
coverage of the Fourier domain ({\em uv} coverage, expressed in units
of spatial frequency defined as a ratio of instantaneous projected
baseline length to the observing wavelength, $\lambda$) of the data
recorded for 0642$+$449.

\subsection{Data correlation}

The data correlation was performed at the DiFX software correlator of
the MPIfR. Fringe search for {\em RadioAstron} was performed for every
scan, and ad-hoc values for delay offsets and rates were determined
and applied, in order to compensate for the acceleration terms of
the spacecraft. The initial fringe-search was performed with 1024
  spectral channels per IF and an integration time of 0.1 sec
  (providing respective search ranges of 10 microseconds and 10$^{-9}$
  sec/sec for the delay and the delay rate), in order to accommodate
for potential large residual delays at the SRT. Fringes between the
SRT and the reference antenna of Effelsberg were found with a residual
delay offset of only about 1 microsecond or less, and a delay rate of
about 2$\times$10$^{-11}$ sec/sec (60 cm/s). The fringe solutions
for the SRT were typically found at high signal-to-noise ratio (SNR):
all but three scans showed fringes with SNR$>$10. For the three scans
with low SNR, values of delay offset and rate were interpolated from
adjacent scans. Final correlation was performed with 32 spectral
channels per IF (with the corresponding channel width of 500 kHz) and
0.5 sec of integration time.

% Figure 3: possm plots
\begin{figure}[ht!]
\centerline{\includegraphics[width=0.45\textwidth,trim=60 90 45 80,clip=true]{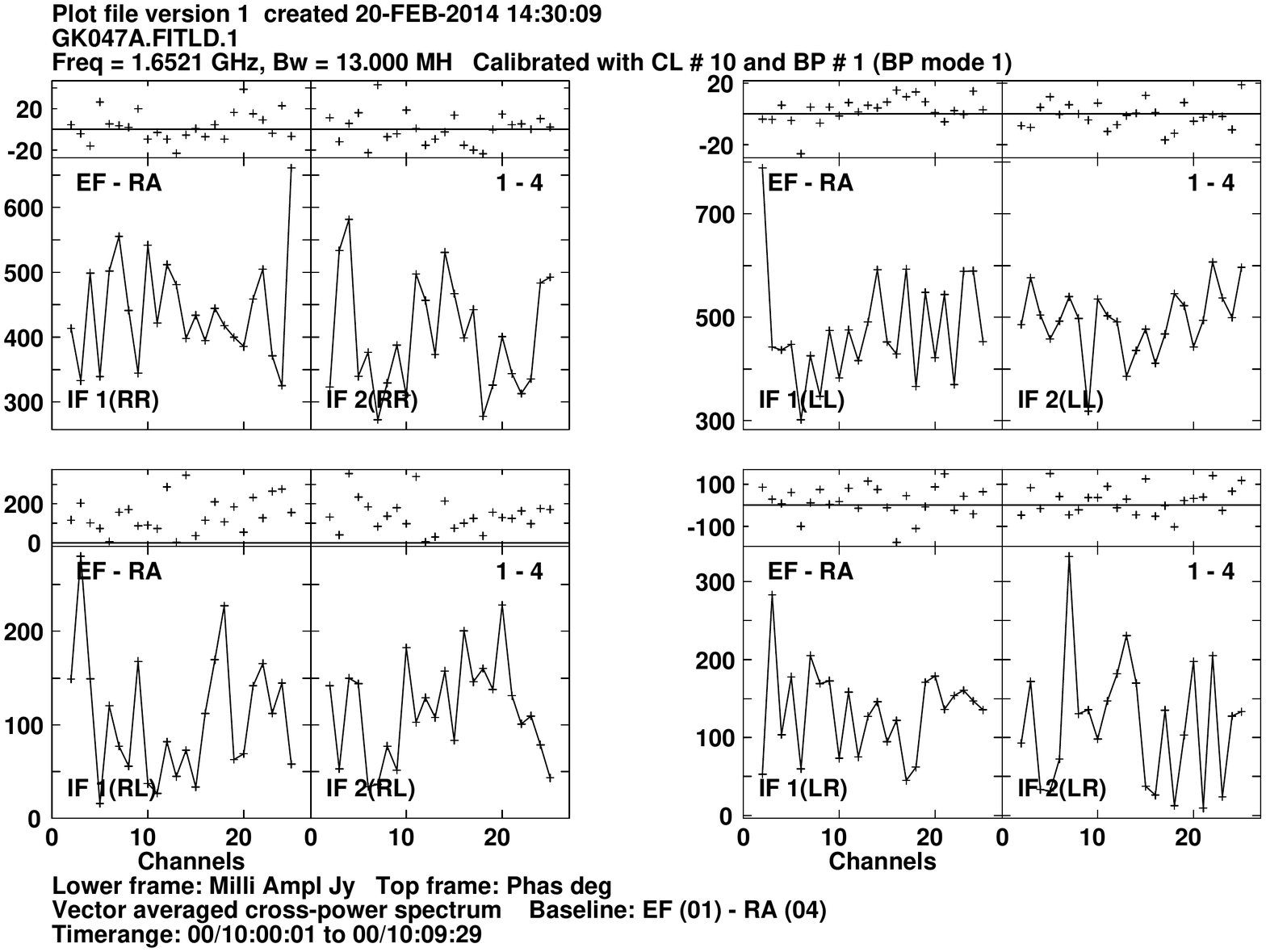}}
\centerline{\includegraphics[width=0.45\textwidth,trim=60 78 45 93,clip=true]{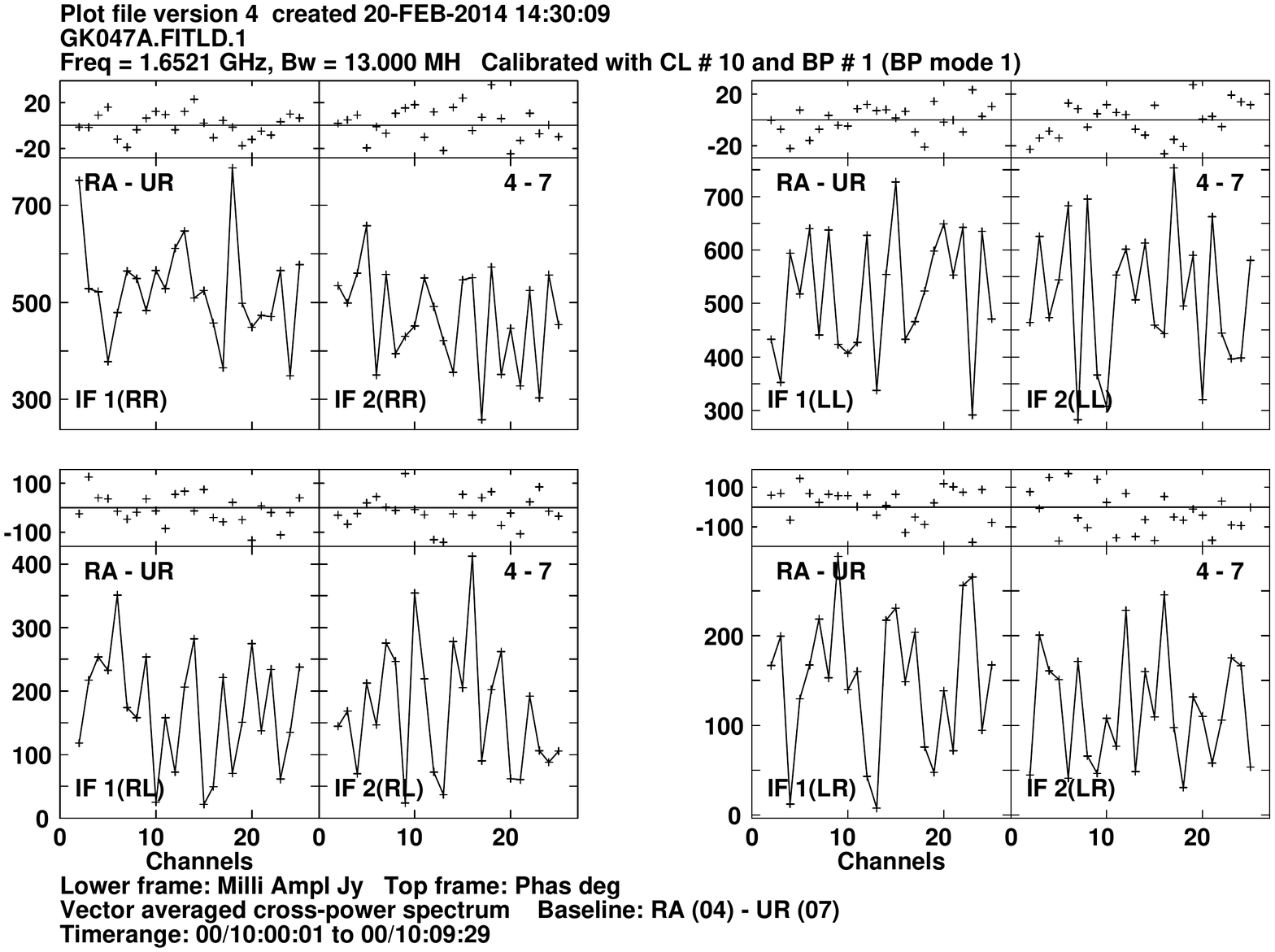}}
%\centerline{\includegraphics[height=0.4\textwidth,angle=-90,clip=true]{ra-bas-p1.ps}}
%\centerline{  \includegraphics[height=0.4\textwidth,angle=-90,clip=true]{ra-bas-p4.ps}}
  \caption{Bandpass amplitudes and phases on ground-space baselines
    between the SRT and Effelsberg (top two rows) and Urumqi (bottom
    two rows), after applying global fringe fitting solutions. The
    bandpasses for total intensity (parallel hands: LL, RR) and
    polarized emission (cross-hands: LR, RL) are shown. }
\label{fg:fringe-tot} 
\end{figure}
%l b r t

% Figure 4: radplot
\begin{figure*}[ht!]
  \centering
\includegraphics[width=0.9\textwidth,angle=0,clip=true]{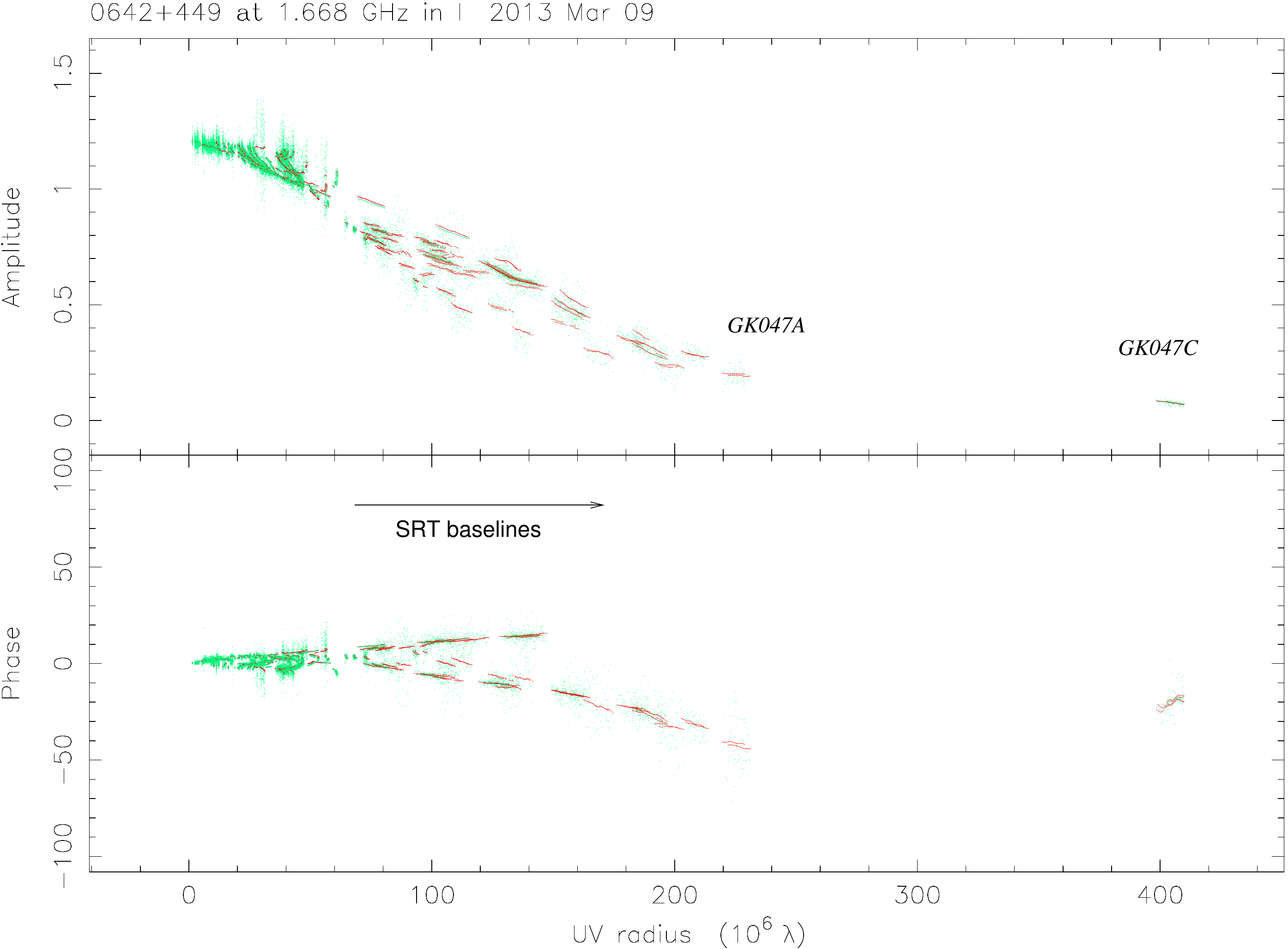}
  \caption{Visibility amplitude (top) and phase (bottom) distributions
    as a function of {\em uv} radius, overplotted with the CLEAN
    model (red color) obtained during the hybrid imaging of the source
    structure. }
\label{fg:0642-radplot} 
\end{figure*}
%The inset in the top panel shows the visibility
%amplitudes and the CLEAN model obtained for the longest baselines
%of the observation.

\subsection{Post-correlation data reduction}

The correlated data were reduced in several steps using
AIPS\footnote{Astronomical Image Processing Software of the National
  Radio Astronomy Observatory, USA} and {\em Difmap}
\citep{shepherd1997,shepherd2011}. The initial calibration was
performed in AIPS, imaging was done using {\em Difmap}, and both packages
were used for calibrating the instrumental polarization and providing
absolute calibration of the polarization vectors. The segments GK047A
and GK047C were calibrated independently and combined together for the
final imaging. Polarization calibration was performed using the
segment GK047A (the
  parallactic angle coverage of GK047C was insufficient for
  using this segment for the polarization calibration).

The {\em a priori} amplitude calibration was applied using the
default antenna gains and the system temperature measurements made at
each antenna during the observation. For the SRT, the sensitivity
parameters measured during the IOC \citep{kovalev+2014} were used. The
data were edited using the flagging information from the station log
data. Parallactic angle correction was applied, to correct for feed
rotation with respect to the target sources. 

\subsubsection{Fringe fitting}

The data were fringe fitted in two steps, first by applying manual
phase-cal corrections and then using the global fringe search for
antenna rates and single and multi-band delays. The group delay
difference between the two polarization channels was corrected. The
resulting post-fringe residual delay rates obtained for the SRT are
plotted in Fig.~\ref{fg:fringe}. These slowly evolving residuals agree well with the
expected accuracy of the orbital velocity determination for the SRT
\citep{kardashev+2013}, but they still should be viewed only as an
indicator of the fidelity of the fringe fitted data, while a more
detailed discussion of the accuracy of the orbit determination for the
SRT is presented elsewhere \citep{duev+2015}.

\subsubsection{Bandpass calibration}

After the fringe fitting,
the receiver bandpasses were corrected. The bandpass-calibrated data
on the baselines between the SRT and the ground antennas in Effelsberg
and Urumqi is shown in Fig.~\ref{fg:fringe-tot}.  The interferometric
visibility signal is clearly detected both in the total intensity (LL
and RR) and the cross-polarization (LR and RL) channels, demonstrating
the excellent quality of the data. Table~\ref{tb:baserrors} lists
typical amplitude and phase errors of the visibility data on the
longest baselines to the SRT observed in the segment GK047A,
demonstrating that the source was detected on baselines to all
participating antennas (the errors were calculated by vector averaging
the calibrated data over two IFs and over a scan length of 10 minutes,
and hence these errors are conservative estimates containing also
contributions from residual bandpasses and offsets between the IFs,
and from amplitude and phase variations over the scan time).

% Table 2
\begin{table}[ht!]
\caption{Typical amplitude and phase errors on the baselines to the SRT}
\label{tb:baserrors}
\begin{center}
\begin{tabular}{l|rr|rr}\hline\hline
Ant. & \multicolumn{4}{|c}{Polarization} \\ \cline{2-5}
Code & \multicolumn{2}{|c|}{LL,RR} & \multicolumn{2}{c}{LR,RL} \\ \cline{2-5}
     & \mc{$\sigma_\mathrm{amp}$} & \multicolumn{1}{c|}{$\sigma_\mathrm{ph}$} & \mc{$\sigma_\mathrm{amp}$} & \mc{$\sigma_\mathrm{ph}$} \\ 
     &  [\%] & [$^{\circ}$] &  [\%] & [$^{\circ}$] \\ \hline
EF   &  5.5 & 2.9 & 22.6 & 50.2 \\
JB   &  9.3 & 3.0 & 30.7 & 51.9 \\
WB   &  6.0 & 3.1 & 30.5 & 46.8 \\
TR   & 11.6 & 4.6 & 38.3 & 54.2 \\
UR   & 13.6 & 6.5 & 45.0 & 60.5 \\
SH   & 53.9 & 56.6& 53.9 & 84.6 \\
NT   & 18.3 & 9.3 & 54.1 & 80.0 \\
HH   & 52.8 & 65.1& 50.8 & 67.2 \\
GB   &  6.5 & 2.6 & 33.4 & 42.7 \\
ZC   & 25.1 & 13.8& 50.6 & 57.7 \\ \hline
\end{tabular}
\end{center} {\bf Notes:}~Columns represent average fractional
amplitude, $\sigma_\mathrm{amp}$ and phase, $\sigma_\mathrm{ph}$,
errors of the total (LL, RR) and cross-polarized (LR, RL) correlated
flux density detected on the SRT baselines to the ground radio
telescopes. 
\end{table}

% Table 3
\begin{table}[h!]
\caption{Instrumental polarization (D-terms)}
\label{tb:dterms}
\begin{center}
\begin{tabular}{l|rr|rr}\hline\hline
Antenna & \multicolumn{2}{|c|}{RCP} &\multicolumn{2}{c}{LCP} \\ \cline{2-5} 
        & $m$ & $\chi$ & $m$ & $\chi$ \\
        & [\%]   & [$^\circ$] & [\%] & [$^\circ$] \\ \hline
RA      &  6.8$\pm$0.3 &  82$\pm$6 &  8.2$\pm$0.8 &   98$\pm$1 \\
        &  7.0$\pm$0.2 &  81$\pm$8 &  8.7$\pm$0.9 &   98$\pm$2 \\
EF      &  2.6$\pm$0.3 &  10$\pm$2 &  2.3$\pm$0.1 &  167$\pm$9 \\
        &  3.1$\pm$0.3 &   8$\pm$3 &  2.4$\pm$0.1 &  185$\pm$9 \\
JB      &  1.3$\pm$0.4 & 140$\pm$7 &  3.5$\pm$0.6 &  336$\pm$3 \\
        &  0.8$\pm$0.4 & 306$\pm$7 &  4.3$\pm$0.6 &  330$\pm$3 \\
WB      &  4.6$\pm$0.5 & 176$\pm$3 &  2.5$\pm$0.2 &  166$\pm$13 \\
        &  4.6$\pm$0.9 & 182$\pm$3 &  2.4$\pm$0.4 &  199$\pm$12 \\
TR      &  8.0$\pm$0.5 & 160$\pm$3 &  7.2$\pm$0.6 &   14$\pm$8  \\
        &  8.7$\pm$0.4 & 164$\pm$3 &  7.6$\pm$0.3 &   31$\pm$8  \\
UR      & 11.6$\pm$1.0 & 271$\pm$12& 10.2$\pm$1.1 &   56$\pm$18 \\
        & 12.5$\pm$1.1 & 294$\pm$12& 12.2$\pm$1.2 &   93$\pm$18 \\
SH      &  6.3$\pm$0.5 & 173$\pm$31&  5.5$\pm$0.4 &   13$\pm$30 \\
        &  6.3$\pm$1.0 & 170$\pm$32&  4.5$\pm$0.8 &  349$\pm$30 \\
NT      &  5.8$\pm$0.3 & 247$\pm$5 &  5.9$\pm$0.2 &  239$\pm$6 \\
        &  5.3$\pm$0.2 & 184$\pm$5 &  5.4$\pm$0.2 &  178$\pm$8 \\
HH      & 21.7$\pm$1.8 & 151$\pm$23& 19.9$\pm$5.2 &   80$\pm$35 \\
        & 18.1$\pm$1.8 & 198$\pm$23&  9.5$\pm$5.2 &  149$\pm$35 \\
GB      &  5.4$\pm$0.3 & 248$\pm$6 &  5.1$\pm$0.8 &  295$\pm$10 \\
        &  4.8$\pm$0.3 & 236$\pm$6 &  3.5$\pm$0.8 &  315$\pm$10 \\
ZC      &  8.0$\pm$0.7 & 160$\pm$7 &  5.2$\pm$1.0 &  146$\pm$3 \\
        &  9.5$\pm$0.7 & 146$\pm$7 &  7.2$\pm$1.0 &  141$\pm$3 \\\hline
\end{tabular}
\end{center} {\bf Notes:}~For each antenna, listed are the fractional
amplitude, $m$, and phase, $\chi$, of the instrumental polarization
(polarization leakage) in the right (RCP) and left (LCP) circular
polarization channel and in the first (IF1, top row) and second (IF2,
bottom row) intermediate frequency channel. The D-term phases have 
been corrected for the phase offsets obtained from the EVPA calibration described in Sect.~\ref{sc:evpa}.
\end{table}

\subsubsection{Polarization calibration}

After applying the bandpass calibration, the calibrated Stokes I data
were exported to {\em Difmap} and imaged using hybrid imaging with
CLEAN deconvolution and self-calibration (with full phase
self-calibration and limited amplitude calibration using a single
amplitude gain correction per antenna). The resulting self-calibrated
visibility amplitudes and phases (and their CLEAN model
representations) are shown in Fig.~\ref{fg:0642-radplot} for all of
the baselines as a function of Fourier spacing ({\em uv}
distance). The visibilities indicate that the source structure is
clearly detected up to the longest ground-space baselines of the
observation.

The polarization calibration proceeded then with the self-calibrated
model of the source structure obtained. The model was
imported into AIPS and used as the calibration model for the full
Stokes data, with amplitude and phase calibration applied on time
intervals of 1 minute.

%\subsubsection{Instrumental polarization}
% Radioastron D-terms.

The instrumental polarization (polarization leakage terms or {\em
  D-terms}, comprising the fractional amplitude and phase) was
obtained using the LPCAL method \citep{leppanen+1995}, with Effelsberg
used as the reference antenna. The resulting polarization leakages
determined through this procedure are listed in
Table~\ref{tb:dterms}. Errors of the D-terms were estimated from
averaging the respective values determined for each of the calibrators
and for the target itself. To estimate the errors for the SRT (which
did not observe the calibrators), the D-terms obtained from each of
the calibrators with the ground antennas were applied to the data on
0642$+$449 and respective solutions for the D-terms of the SRT were
obtained and compared with the solution obtained directly from the
target data (hence the D-terms errors for the SRT might be
underestimated). The amplitude of instrumental polarization of the SRT
is found to be within 9\% and with excellent phase consistency between
the IFs. This figure cannot be directly compared with the results
  from pre-launch measurements, as those measurements were made only
  for the antenna feeds of the SRT, yielding an $\approx 2$\% leakage at 1.6
  GHz \citep{turygin2014}. The polarization leakage of the SRT is
comparable to the instrumental polarization obtained for the ground
antennas, demonstrating robust polarization performance of the SRT and
ensuring reliable polarization measurements with data on all
ground-space baselines. Similar estimates of the L-band polarization
leakage were obtained for the SRT from the analysis of the {\em
  RadioAstron} AGN survey observations \citep{pashchenko+2015}.

% Table 4
\begin{table}[t!]
\caption{Effelsberg measurements of total and polarized flux densities of the targets}
\label{tb:efpol} 
\begin{center}
\begin{tabular}{c|rrr}\hline\hline
Source   & \mc{$S_\mathrm{tot}$} & \mc{$S_\mathrm{pol}$}& \mc{$\chi$} \\
         & \mc{[Jy]}       & \mc{[Jy]}       & \mc{[deg]}    \\\hline
0642+449 & $~1.32\pm 0.02$ & $0.022\pm 0.008$& $77.2\pm 12.0$ \\
0851+202 & $~2.77\pm 0.03$ & $0.069\pm 0.016$& $32.1\pm ~~1.7$ \\
1328+307 & $13.68\pm 0.21$ & $1.37\pm 0.10~~$& $33.3\pm ~~1.4$ \\\hline
\end{tabular}
\end{center} {\bf Notes:}~Measurements were made on 9-10 March
2013. Column designation: $S_\mathrm{tot}$ -- total flux density;
$S_\mathrm{pol}$ -- polarized flux density; $m$ -- percentage of
polarization; $\chi$ -- position angle of the electric vector position angle (EVPA).
\end{table}

% Figure 5: calibrator images
\begin{figure}[b!]
  \centering
%\centerline{
% \includegraphics[width=0.25\textwidth]{1328+307_polimage.eps}
% \includegraphics[width=0.25\textwidth]{0851+202_polimage.eps}
%}
\centerline{
 \includegraphics[width=0.25\textwidth]{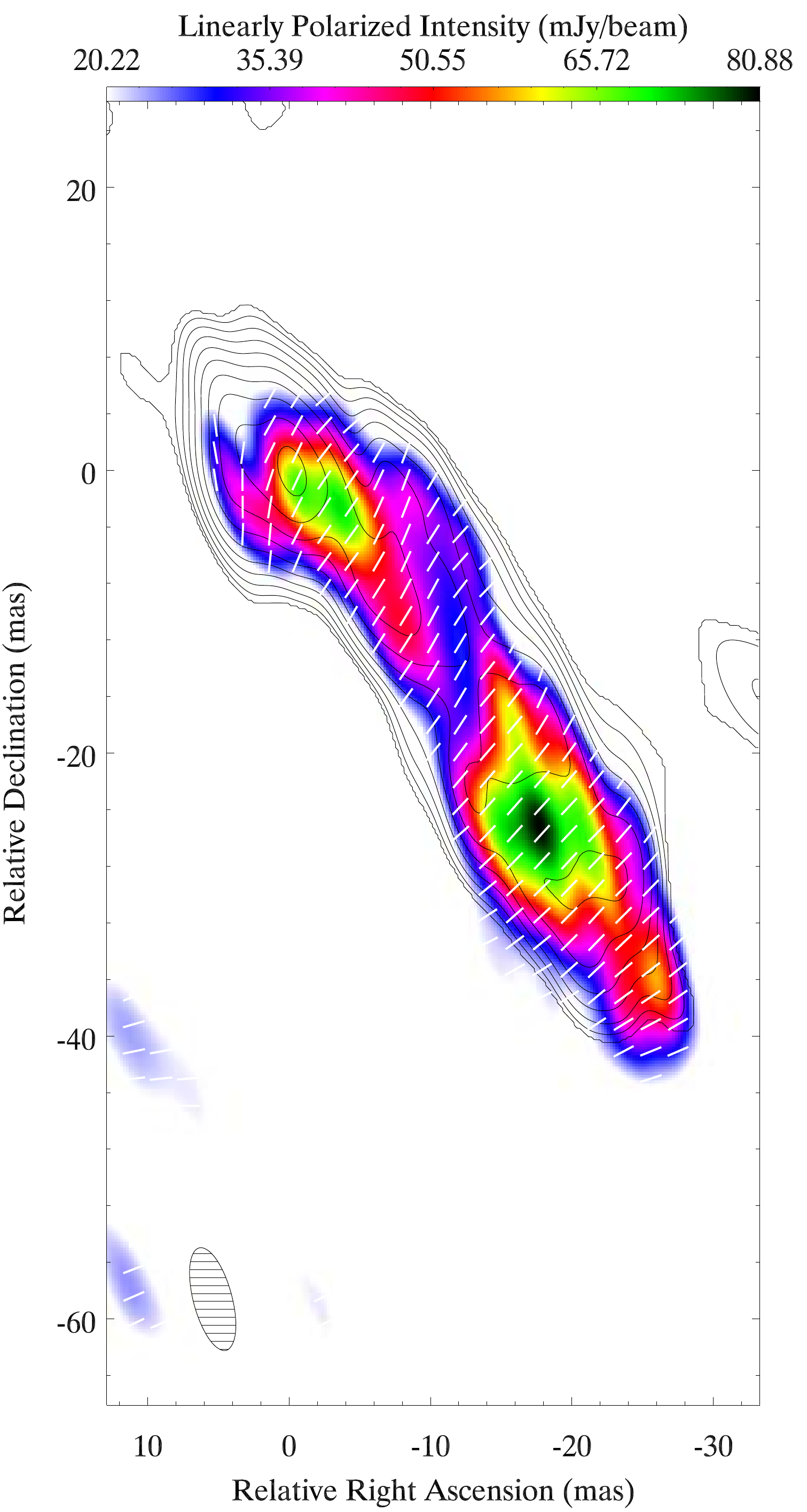}
 \includegraphics[width=0.25\textwidth]{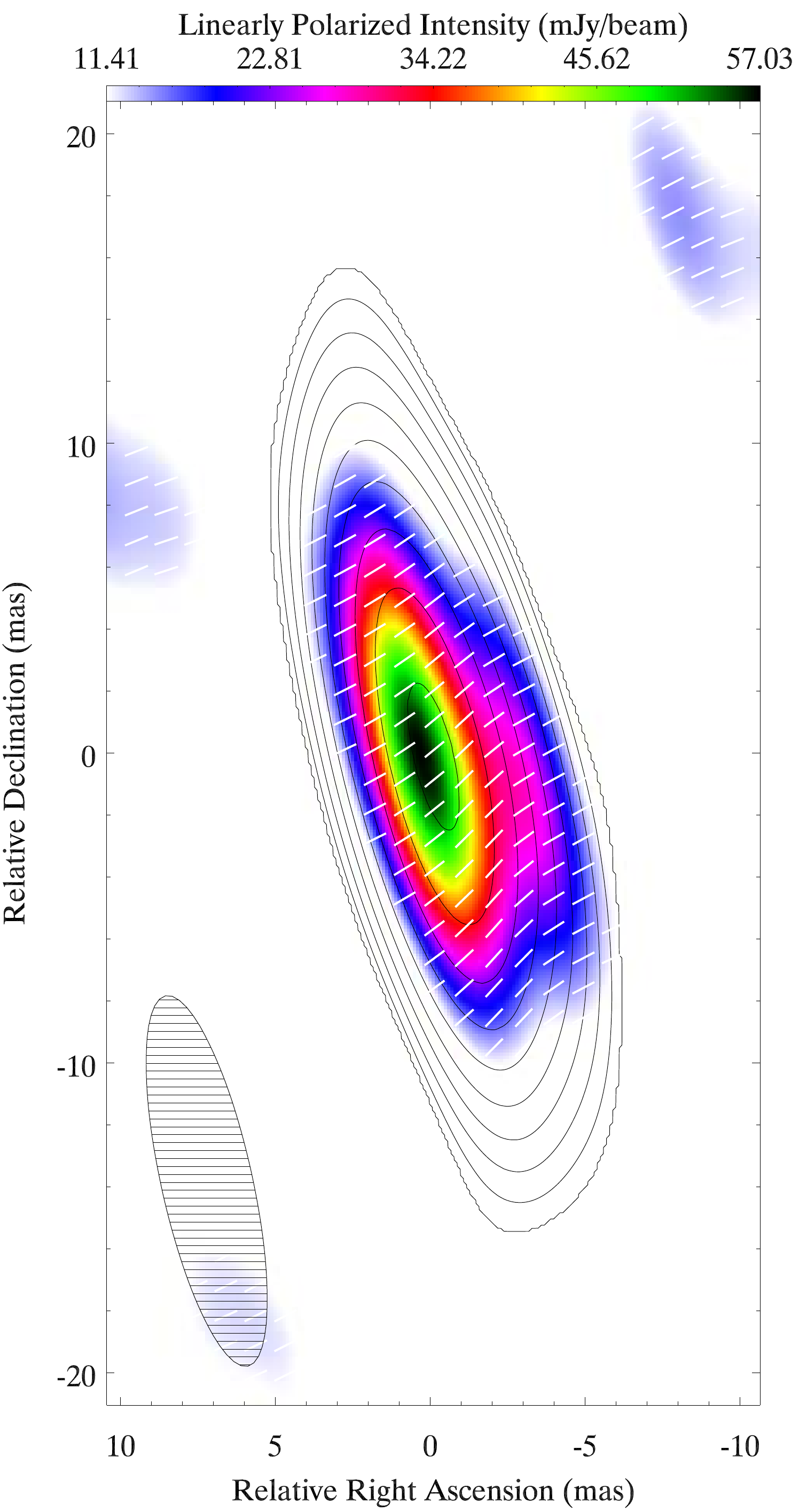}
}
\caption{Ground array images of the calibrator sources 1328$+$307
  (left) and 0851$+$202 (right), made using the uniform data weighting. Image parameters are listed in
  Table~\ref{tb:mapspar}. Contours show the total intensity; colors
  indicate linearly polarized intensity; vectors show the EVPA
  orientation.}
\label{fg:calmaps}
\end{figure}

% Figure 6: image of 0642
\begin{figure*}[ht!]
  \centering
  \includegraphics[width=0.9\textwidth]{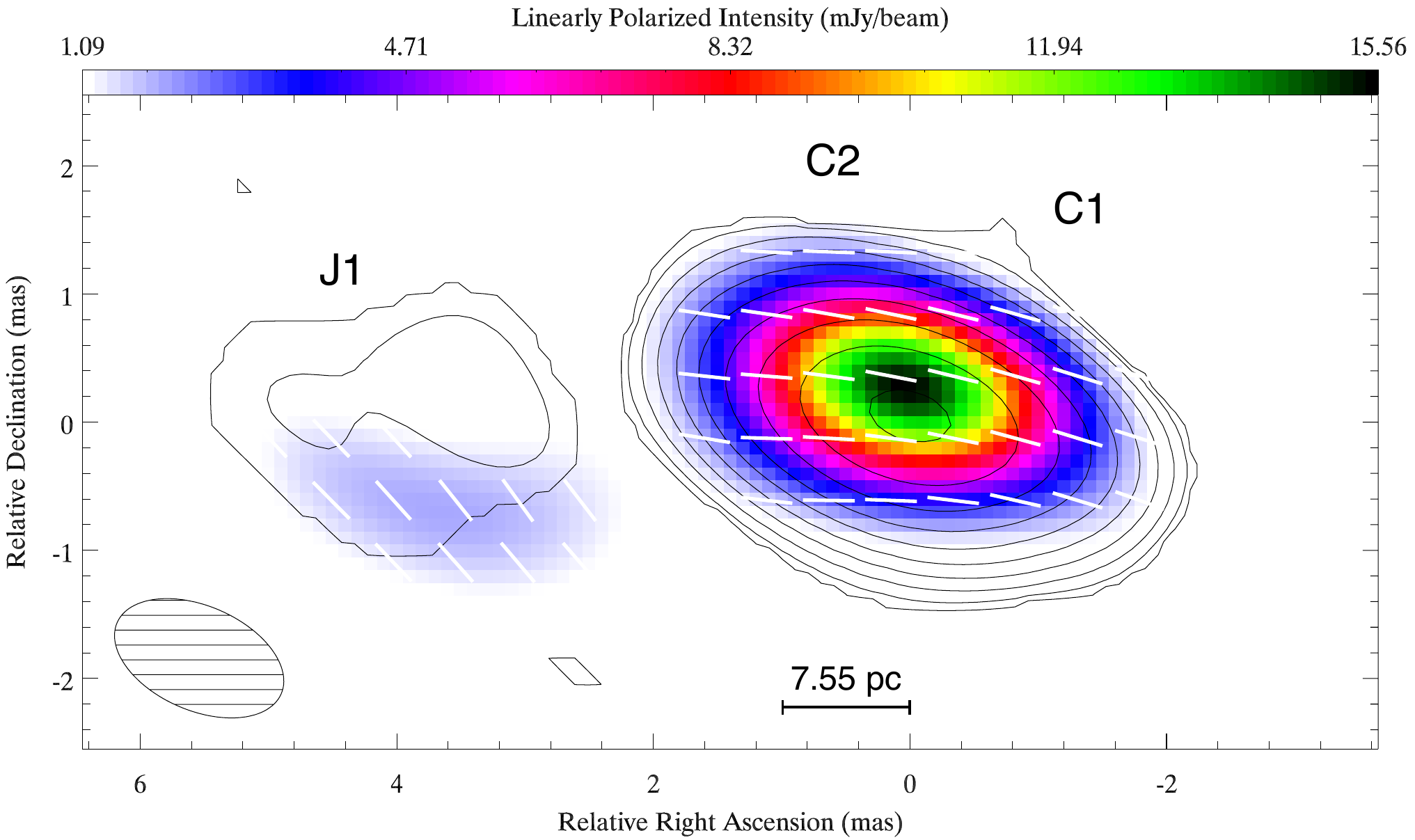}
  \caption{{\em RadioAstron} space VLBI image of 0642$+$449 of total
    (contours) and linearly polarized (color shades) emission,
    made using the uniform data weighting. The orientation of
    linear polarization is indicated by vectors. Basic parameters of
    the total intensity image are listed in Table~\ref{tb:mapspar}. Labels indicate the modelfit decomposition of the source structure described in Table~\ref{tb:modelfit}.}
\label{fg:0642-map} 
\end{figure*}

After applying the D-terms determined for the SRT and the ground
antennas, the calibrated full Stokes data were exported into {\em Difmap}, imaged in Stokes I, Q, and U, and combined together,
providing images of the total intensity, linearly polarized intensity,
and polarization vectors (electric vector position angle, EVPA).

\subsubsection{EVPA calibration}
\label{sc:evpa}

The absolute EVPA calibration was performed using single-dish
measurements of the total flux density polarization of the target and
calibrator sources. These measurements (listed in
Table~\ref{tb:efpol}) were made concurrently with the {\em RadioAstron}
observation at the Effelsberg 100-meter telescope. The averaged EVPA
of the polarized emission detected in the VLBI data was related to the
polarization vectors that were measured with the Effelsberg antenna,
and the resulting EVPA offsets were applied to the VLBI images. The
resulting ground array images of the calibrator sources are shown in
Fig.~\ref{fg:calmaps}, and the image parameters are listed in
Table~\ref{tb:mapspar}.

%Mean epohc: JD~2456361.3
%Frequency 1.66E9

%zzz

\section{Polarization image of 0642$+$449}
\label{sc:image}

The {\em RadioAstron} polarization image of 0642$+$449 is shown in
Fig.~\ref{fg:0642-map}, and the basic parameters of the image are
listed in Table~\ref{tb:mapspar}. The source shows a compact
core-dominated structure, with a resolved nuclear region which
contains about 98\% of the total flux density in the image. The weak
and extended jet feature, J1, located at about 4 mas separation from
the nucleus has a peak brightness of $\approx 15$\,mJy/beam, which is
not surprising considering energy losses due to adiabatic
  expansion of the flow up to the estimated deprojected linear
separation of 2.3\,kpc of this feature from the core of the source
(assuming the jet viewing angle of 0.8$^\circ$;
\citeauthor{pushkarev+2009} \citeyear{pushkarev+2009}). The linear
  separation can be a factor of $\sim 3.5$ smaller, if the observed
  misalignment of $\approx 2^\circ$ between the position angles of C2
  and J1 reflects the physical change of the jet direction.  The
  observed concentration of the polarized flux density along the
  southern edge of this feature may reflect interaction with the
  ambient medium occurring in this region, although it may also result
  from the errors of the D-terms determination or from other
  instrumental effects \citep[cf.][]{gabuzda+2001,hovatta+2012}.

% Table 5
\begin{table}[b!]
\caption{Parameters of total intensity images}
\label{tb:mapspar}
\begin{center}
\begin{tabular}{r|ccccc}\hline\hline
Source & $S_\mathrm{tot}$ & $S_\mathrm{peak}$ & $S_\mathrm{neg}$ & $\sigma_\mathrm{rms}$ & Beam \\ \hline
%  & [Jy] & [Jy/beam] & [mJy/beam] & [mJy/beam] & [mas,mas,deg] \\ \hline
0851$+$202 & 2.06 & 1.73 &-19.0 & 6.5  & 12.2,2.93,12.6 \\ 
1328$+$307 & 7.87 & 0.89 &-47.1 & 8.6 & 7.48,2.73,14.9 \\ 
0642$+$449 & 1.21 & 0.81 & 3.6 & 0.7  & 1.38,0.81,66.5 \\ \hline
\end{tabular}
\end{center}
{\bf Notes:}~Column designation: $S_\mathrm{tot}$\,[Jy] -- total flux density; $S_\mathrm{peak}$\,[Jy/beam] -- peak flux density; $S_\mathrm{neg}$ [mJy/beam] -- maximum negative flux density in the image; $\sigma_\mathrm{rms}$\,[mJy/beam] -- rms noise in the image; Beam: major axis, minor axis, position angle of major axis [mas,\,mas,\,$^{\circ}$].
\end{table}

% Table 6
\begin{table*}[ht!]
\caption{Decomposition of compact structure in 0642$+$449}
\label{tb:modelfit}
\begin{center}
\begin{tabular}{lrrrccrccc}\hline\hline
Comp. & \mc{$S_\mathrm{tot}$} & \mc{$r$} & \mc{$\phi$} & $\theta$ & 
$\theta_\mathrm{min}$ & \mc{$T_\mathrm{b}$} & $S_\mathrm{pol}$ & $m_\mathrm{pol}$ & $\chi_\mathrm{pol}$ \\
      & \mc{[mJy]} & \mc{[mas]} & \mc{[$^{\circ}$]} & \mc{[mas]} & \mc{[mas]} & \mc{[10$^{11}$\,K]} & [mJy] & [\%] & [$^{\circ}$]  \\ \hline
C1    & 462$\pm$16 & 0.38$\pm$0.01 & $-$88.3$\pm$0.7  & 0.40$\pm$0.01  & 0.16  & 12.8$\pm$0.6 & 7.7$\pm$0.8 & 1.7$\pm$0.2 & 79$\pm$3\\
C2    & 721$\pm$19 & 0.32$\pm$0.01  & 70.1$\pm$1.1  & 0.60$\pm$0.01  & 0.13  &  8.8$\pm$0.3 & 9.9$\pm$0.7 & 1.4$\pm$0.1 & 89$\pm$2 \\
J1    & 20$\pm$9  & 3.80$\pm$0.20  & 92.0$\pm$3.2 & 0.89$\pm$0.43  & 0.75  &  0.11$\pm$0.04 & 1.7$\pm$0.7 & 8.5$\pm$5.2 & 39$\pm$11 \\ \hline
\end{tabular}
\end{center} {\bf Notes:}~Gaussian model description: $S_\mathrm{tot}$
-- total flux density; component position in polar coordinates
($r,\,\phi$) with respect to the map center; component size, $\theta$;
minimum resolvable size, $\theta_\mathrm{min}$, of the respective
component; brightness temperature, $T_\mathrm{b}$, in the observer's
frame, derived for the parameters of the Gaussian fit. Polarization properties (measured from the Stokes U and Q images: $S_\mathrm{pol}$ -- flux density of linearly polarized emission; $m_\mathrm{pol}$ -- fractional polarization; $\chi_\mathrm{pol}$ -- polarization position angle in the sky plane (measured North-through-East).
\end{table*}

The core region is clearly resolved and extended along a P.A. of
$\approx 81^{\circ}$, It can be decomposed into two circular Gaussian
components. The resulting three-component decomposition of the source
structure is presented in Table~\ref{tb:modelfit}. The two nuclear
components, C1 and C2, are separated by 0.76\,mas ($\approx 410$\,pc,
deprojected). Similar structural composition of the nuclear region is
also detected in the ground VLBI from the MOJAVE survey at 15\,GHz
\citep{lister+2013}. The same structure is also present in the images
of 0642$+$449 made at 24 and 43\,GHz \citep{fey+2000}.

Overall, the structure observed in 0642$+$449 may be similar to the
main features of the jet in M\,87 \citep[cf.][]{sparks+1996}, in which
a prominent standing shock (component HST-1) is located at
200--280\,pc (considering the uncertainty of the jet viewing angle)
and a prominent feature A located at 2.8--3.7\,kpc distance from the
core is believed either to be an oblique shock \citep{bicknell+1996}
or to result from the helical mode of Kelvin-Helmholtz instability
\citep{lobanov+2003}. This may also explain the apparent displacement
between the total intensity and linearly polarized emission in the
region J1, which is similar to the polarization asymmetries observed in
M\,87 at these linear scales \citep{owen+1989}.

The polarized emission detected in 0642$+$449 is dominated
by the two nuclear components, with fractional polarization
reaching 15.6\,mJy ($\approx 2$\%), and the magnetic field showing
predominantly transverse orientation with respect to the general jet
direction (assuming that the innermost component C1 corresponds to the
canonical ``VLBI core'' region in which the emission becomes optically
thin at the frequency of observation; \citeauthor{koenigl1981}
\citeyear{koenigl1981}). The polarization vectors vary by about
$10^{\circ}$ across the nucleus, with the polarization angle observed
in the most compact region C1 ($\approx 80^{\circ}$) coming close to
the polarization angle measured with Effelsberg. A substantial
rotation of the EVPA vectors is seen in some of the MOJAVE images of
0642$+$449, which is likely to result from plasma propagation in a jet
observed at an extremely small viewing angle. The weak ($\sim
2$\,mJy/beam) linearly polarized emission in the jet component J1
corresponds to a fractional polarization of $\sim 15$\%, with a large
error margin on this estimate. The magnetic field is roughly
longitudinal in this region.

%\subsection{Rotation measure}

The MOJAVE database\footnote{http://www.physics.purdue.edu/MOJAVE/}
lists, for the frequency of 15\,GHz, an EVPA of 122$^\circ$ in March
2011. Analysis of the multifrequency MOJAVE observations
\citep{hovatta+2012} yields, for 0642$+$449, a median rotation measure
of $-268$\,rad/m$^2$ over the whole source structure, with a slightly
higher rotation measure of $-280$\,rad/m$^2$ over the compact core
region.

Assuming a linear trend in the EVPA in time as derived from the two
latest MOJAVE observations, the polarization angle should be $\sim
105^{\circ}$ in March 2013 at 15\,GHz. We measure $\chi =
80^{\circ}$--90$^{\circ}$ over the core region. Formally, this
corresponds to a rotation measure of $-15$\,rad/m$^2$, under
assumption that the rotation measure is not affected by the core
shift. It would then require about three turns of phase to get this
value to be reconciled with the RM reported from the MOJAVE
measurements. Alternatively, the smaller RM estimated from the {\em
  RadioAstron} measurements may reflect the reported trend of
decreasing RM with decreasing frequency \citep{kravchenko+2015}. It
would still require about one turn of phase to bring the estimated RM
value to a better agreement with the empirical dependence
$|\mathrm{RM}|(\nu)\propto \nu^{1.6}$ derived in
\cite{kravchenko+2015}.

% Figure 7
\begin{figure*}[ht!]
  \centering
  \includegraphics[width=0.9\textwidth]{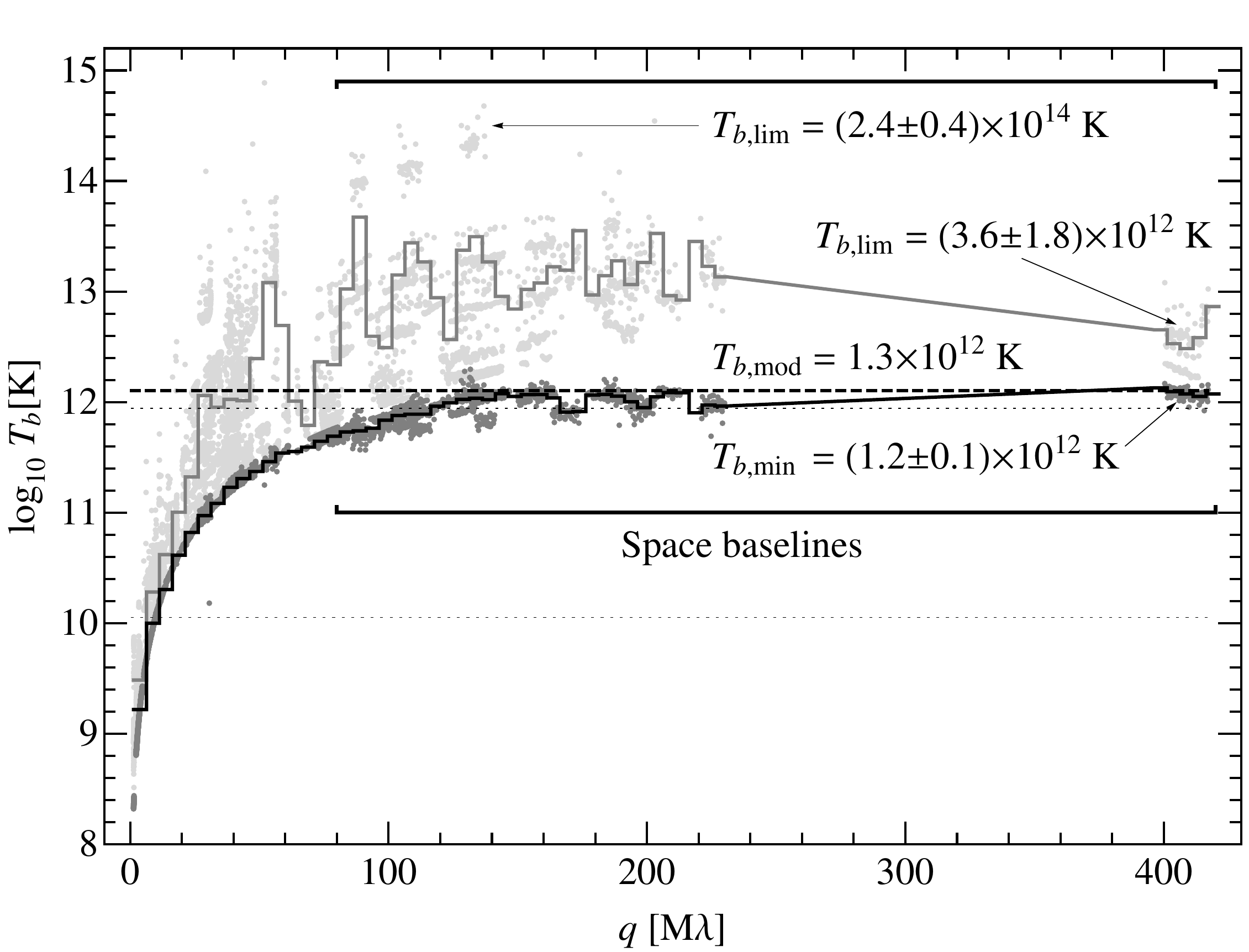}
  \caption{Visibility-based estimates of the brightness temperature in
    0642$+$449, calculated in the observer's frame. Data points
    represent the minimum (dark gray), $T_\mathrm{b,min}$, and maximum
    limiting (light gray), $T_\mathrm{b,lim}$, brightness temperatures
    derived from individual visibilities measured at different {\em
      uv} distances, $q$ \citep{lobanov2015}. The two-dimensional
    distribution of $T_\mathrm{b,lim}$ is also presented in
    Fig.~\ref{fg:0642-uvplot} with a color wedge. Braces above and below the
    brightness temperature data indicate the range of {\em uv}
    distances covered by the space baselines of the {\em RadioAstron}
    observation. The histograms show the respective brightness
    temperatures averaged over bins of 5\,M$\lambda$ in size. The
    thick dashed line indicates the brightness temperature,
    $T_\mathrm{b,mod} = 1.3\times 10^{12}$\,K, derived for the most
    compact feature, C1, of the Gaussian fit of the source structure
    described in Table~\ref{tb:modelfit}. The dotted lines indicate
    the respective brightness temperatures of the other two Gaussian
    components. The minimum brightness temperature, $T_\mathrm{b,min}
    = 1.2 \times 10^{12}$\,K, is constrained by the visibilities with
    $q>380$\,M$\lambda$. The maximum limiting brightness temperature
    $T_\mathrm{b,lim} = 2.4 \times 10^{14}$\,K can be estimated from
    the data on the SRT baselines to Noto and Hartebeesthoek,
    although potential problems with amplitude calibration cannot be
    ruled out. A more conservative estimate of $T_\mathrm{b,lim} =
    3.6\times 10^{12}$\,K can be obtained from the data on the longest
    space-ground baselines.}
\label{fg:0642-tb} 
\end{figure*}

\subsection{Brightness temperature}
\label{sc:tb}

The Gaussian modelfit described in Table~\ref{tb:modelfit} can be used
for obtaining estimates of brightness temperature (also presented in
the same Table), with the highest brightness temperature of $1.3\times
10^{12}$\,K associated with the most compact core feature C1. This
value is somewhat lower than the one that should be expected at
1.6\,GHz, based on the trend $\log_{10}T_\mathrm{b}[\mathrm{K}] = 13.3 -
1.1\,\log_{10}\nu[\mathrm{GHz}]$ that can be obtained from the existing
measurements at 5, 15, and 86\, GHz
\citep{lobanov+2000,kovalev+2005,dodson+2008,lee+2008}. However, since the
morphology of the emitting region can be different from a simple
Gaussian shape, the actual brightness temperature of the emission may
differ from the modelfit-based estimates, for instance, if the brightness
distribution is smooth and the respective brightness temperature is
largely determined mostly by the transverse dimension of the flow \citep{lobanov2015}.  To
investigate this possibility, the modelfit-based estimates of
$T_\mathrm{b}$ can be compared with estimates obtained directly from
visibility amplitudes and their errors, which yields the absolute
minimum brightness temperature, $T_\mathrm{b,min}$, and an estimate of
the limiting brightness temperature, $T_\mathrm{b,lim}$ that can
obtained from the data under the requirements that the structural
detail sampled by the given visibility is resolved
\citep{lobanov2015}. 

These two estimates are compared in Fig.~\ref{fg:0642-tb} with the
brightness temperatures calculated from the Gaussian modelfit. This
comparison shows that the visibility amplitudes on the space baselines
longer that $\approx 140$\,M$\lambda$ require the brightness
temperature to be larger than $10^{12}$\,K, which agrees with
$T_\mathrm{b}$ obtained for the most compact feature C1 of the
Gaussian modelfit. The limiting brightness temperature
$T_\mathrm{b,lim} = 3.6 \times 10^{12}$\,K can be estimated at the
longest space baselines. These visibilities concentrate however only
within a narrow range of position angles ($-10\fdg 2 \pm 0\fdg 1$),
with corresponding structural sensitivity limited to ${\mathrm P.A.}
  \approx 80^{\circ}$, which essentially coincides with direction of
  the inner jet (hence resulting in lower visibility amplitudes and,
  consequently, lower estimates of $T_\mathrm{b,lim}$).

Indeed, some of the $T_\mathrm{b,lim}$ estimated from the visibility
amplitudes at shorter space baselines and different position
angles reach above $10^{14}$\,K, notably with a large spread of the
estimates obtained at similar spatial frequencies.  The observed
spread reflects the difference of signal-to-noise ratios (SNR) in the
measurements on different baselines, and it also may indicate
potential biases due to a priori gain calibration and amplitude
self-calibration. The three particularly outstanding clusters of points
in Fig.~\ref{fg:0642-tb}, which have $T_\mathrm{b,lim}>10^{14}$\,K,
come from visibilities on the baselines to Westerbork, Torun, Noto,
and Hartebeesthoek. 

As the visibility SNR on these baselines is
typically lower than on baselines to Effelsberg or Jodrell Bank (which
do not yield such high $T_\mathrm{b,lim}$ at similar radial distances
and position angles), it is most likely that the exceptionally high
brightness temperature limits are caused by problems with the
amplitude calibration of the antenna gains or specific observing scans
with these antennas.

If the outliers are ignored and only the bulk of the
$T_\mathrm{b,lim}$ estimates are considered in the range of
100--200\,M$\lambda$, the average $T_\mathrm{b,lim}$ increases to
$9.1\times 10^{12}$\,K, and this can be viewed as a viable upper limit
on the brightness temperature of the most compact structure in the jet
of 0642$+$449. This corresponds to brightness temperature of
$4.0\times 10^{13}$\,K in the rest frame of the source. If the
brightness temperature is generally determined by the transverse
dimension of the flow, as suggested by the analysis of the MOJAVE
visibility data \citep{lobanov2015}, the inner jet of 0642$+$449
should have a width of $\approx 0.15$\,mas to satisfy the limiting
brightness temperature estimated above.

\section{Summary}
\label{sc:summary}

The {\em RadioAstron} Early Science observations of the high-redshift
quasar TXS~0642$+$449 have tested the polarization performance of the
space radio telescope (SRT) at 1.6\,GHz ($\lambda = 18$\,cm) and
provided a framework for establishing a set of procedures for
correlation, post-processing and calibration of space VLBI
polarization data. The instrumental polarization of the SRT is found
to be less than 9\%, which enables robust reconstruction of the
polarization signal on the ground-space baselines of {\em RadioAstron}
observations and provides a basis for high-fidelity imaging of
polarized emission.

The total and linearly polarized emission from the quasar 0642$+$449
is detected on the SRT baselines of up to 75,560\,km (5.93 Earth
diameters) in length, corresponding to angular scales of 0.5
milliarcseconds (mas).  The resulting {\em RadioAstron} image of
linearly polarized emission in 0642$+$449 has a resolution of
0.8\,mas, which is $\sim 4$ times better than the resolution of ground
VLBI images at 18\,cm.  The image probes the physical conditions in
the jet on angular scales as small as $\approx 0.2$\,mas (linear scale
of $\approx 1.5$\,pc) taking into account the SNR-driven resolution
limits of the observation.

The structure of total intensity and linearly polarized emission
recovered from the {\em RadioAstron} image of 0642$+$449 suggests that
this object is likely to be a ``canonical'' quasar with a powerful
relativistic jet observed at an extremely small angle of sight and
featuring a bright ``core'' of the jet (component C1) and a likely
recollimation shock (component C2) located at a deprojected distance
of about 400\,pc away from the core. Each of these regions is only
weakly polarized, with fractional polarization not exceeding $\approx
2\%$ and polarization vector implying a predominantly transverse
magnetic field (in agreement with expectations for strong shocks in
jet). The weak and more distant feature J1 observed at a separation of
$\approx 3.8$\,mas ($\approx 2$\,kpc, deprojected) has a much stronger
polarization, possibly exceeding a 10\% level. The dominant magnetic
field components is likely to be poloidal in this region, implying
that it either represents a weak shock or may even result from plasma
instability developing in the jet.

The {\em RadioAstron} data have also been used for making brightness
temperature measurements based on the Gaussian decomposition
(modelfit) of the structure and on estimates made directly from the
visibility data. The most robust estimate provided by these
measurements yields a brightness temperature of $\approx 3.6 \times
10^{12}$\,K for the most compact region in the jet. Further analysis
of the {\em RadioAstron} data indicate that the the brightness temperature
of the radio emission from this region cannot be lower that $\sim
10^{12}$\,K and is not likely to exceed $\approx 9\times 10^{12}$\,K
(corresponding to $\approx 4 \times 10^{13}$\,K in the reference frame
of the source).

\section*{Acknowledgments}

The {\em RadioAstron} project is led by the Astro Space Center of the
Lebedev Physical Institute of the Russian Academy of Sciences and the
Lavochkin Scientific and Production Association under a contract with
the Russian Federal Space Agency, in collaboration with partner
organizations in Russia and other countries. 
This research is based on
observations correlated at the Bonn Correlator, jointly operated by
the Max Planck Institute for Radio Astronomy (MPIfR), and the Federal
Agency for Cartography and Geodesy (BKG).  YYK, MML, KVS, PAV are
supported by the Russian Foundation for Basic Research (RFBR) grant
13-02-12103. KVS is also supported by the RFBR grant 14-02-31789.  JLG
acknowledges support from the Spanish Ministry of Economy and
Competitiveness grant AYA2013-40825-P.  
The European VLBI Network is a
joint facility of European, Chinese, South African and other radio
astronomy institutes funded by their national research councils. The
National Radio Astronomy Observatory is a facility of the National
Science Foundation operated under cooperative agreement by Associated
Universities, Inc.

\bibliographystyle{aa}
\bibliography{ra}

\begin{thebibliography}{56}
\expandafter\ifx\csname natexlab\endcsname\relax\def\natexlab#1{#1}\fi

\bibitem[{{Andreyanov} {et~al.}(2014){Andreyanov}, {Kardashev}, \&
  {Khartov}}]{andreyanov+2014}
{Andreyanov}, V.~V., {Kardashev}, N.~S., \& {Khartov}, V.~V. 2014, Cosmic
  Research, 52, 319

\bibitem[{{Andrianov} {et~al.}(2014){Andrianov}, {Girin}, {Zharov}, {Kostenko},
  {Likhachev}, \& {Shatskaya}}]{andrianov+2014}
{Andrianov}, A.~S., {Girin}, I.~A., {Zharov}, V.~E., {et~al.} 2014, Proc. of
  S.A. Lavochkin Association, 3, 55

\bibitem[{{Arsentev} {et~al.}(1982){Arsentev}, {Berzhatyi}, {Blagov},
  {Gvamichava}, {Danilov}, {Dolgopolov}, {Zakson}, {Katsari}, {Krasnov}, \&
  {Kuznetsov}}]{arsentev+1982}
{Arsentev}, V.~M., {Berzhatyi}, V.~I., {Blagov}, V.~D., {et~al.} 1982,
  Akademiia Nauk SSSR Doklady, 264, 588

\bibitem[{{Bicknell} \& {Begelman}(1996)}]{bicknell+1996}
{Bicknell}, G.~V. \& {Begelman}, M.~C. 1996, \apj, 467, 597

\bibitem[{{Bruni}(2014)}]{bruni2014}
{Bruni}, G. 2014, in COSPAR Meeting, Vol.~40, 40th COSPAR Scientific Assembly.
  Held 2-10 August 2014, in Moscow, Russia, Abstract E1.10-22-14., 418

\bibitem[{{Bruni} {et~al.}(2015){Bruni}, {Anderson}, {Alef}, {Lobanov}, \&
  {Zensus}}]{bruni+2015}
{Bruni}, G., {Anderson}, J.~M., {Alef}, W., {Lobanov}, A.~P., \& {Zensus},
  J.~A. 2015, in Proceedings of Science, Vol. EVN 2014, Proceedings of the 12th
  European VLBI Network Symposium, 119

\bibitem[{{Deller} {et~al.}(2011){Deller}, {Brisken}, {Phillips}, {Morgan},
  {Alef}, {Cappallo}, {Middelberg}, {Romney}, {Rottmann}, {Tingay}, \&
  {Wayth}}]{deller+2011}
{Deller}, A.~T., {Brisken}, W.~F., {Phillips}, C.~J., {et~al.} 2011, \pasp,
  123, 275

\bibitem[{{Deller} {et~al.}(2007){Deller}, {Tingay}, {Bailes}, \&
  {West}}]{deller+2007}
{Deller}, A.~T., {Tingay}, S.~J., {Bailes}, M., \& {West}, C. 2007, \pasp, 119,
  318

\bibitem[{{Dodson} {et~al.}(2008){Dodson}, {Fomalont}, {Wiik}, {Horiuchi},
  {Hirabayashi}, {Edwards}, {Murata}, {Asaki}, {Moellenbrock}, {Scott},
  {Taylor}, {Gurvits}, {Paragi}, {Frey}, {Shen}, {Lovell}, {Tingay}, {Rioja},
  {Fodor}, {Lister}, {Mosoni}, {Coldwell}, {Piner}, \& {Yang}}]{dodson+2008}
{Dodson}, R., {Fomalont}, E.~B., {Wiik}, K., {et~al.} 2008, \apjs, 175, 314

\bibitem[{{Duev} {et~al.}(2012){Duev}, {Molera Calv{\'e}s}, {Pogrebenko},
  {Gurvits}, {Cim{\'o}}, \& {Bocanegra Bahamon}}]{duev+2012}
{Duev}, D.~A., {Molera Calv{\'e}s}, G., {Pogrebenko}, S.~V., {et~al.} 2012,
  \aap, 541, A43

\bibitem[{{Duev} {et~al.}(2015){Duev}, {Zakhvatkin}, {Stepanyants}, {Molera
  Calv{\'e}s}, {Pogrebenko}, {Gurvits}, {Cim{\`o}}, \& {Bocanegra
  Baham{\'o}n}}]{duev+2015}
{Duev}, D.~A., {Zakhvatkin}, M.~V., {Stepanyants}, V.~A., {et~al.} 2015, \aap,
  573, A99

\bibitem[{{Fey} {et~al.}(2000){Fey}, {Boboltz}, {Gaume}, \&
  {Johnston}}]{fey+2000}
{Fey}, A.~L., {Boboltz}, D.~A., {Gaume}, R.~A., \& {Johnston}, K.~J. 2000, in
  International VLBI Service for Geodesy and Astrometry 2000 General Meeting
  Proceedings, ed. N.~R. {Vandenberg} \& K.~D. {Baver}, 285--287

\bibitem[{{Ford} {et~al.}(2014){Ford}, {Anderson}, {Belousov}, {Brandt},
  {Ford}, {Kanevsky}, {Kovalenko}, {Kovalev}, {Maddalena}, {Sergeev},
  {Smirnov}, {Watts}, \& {Weadon}}]{ford+2014}
{Ford}, H.~A., {Anderson}, R., {Belousov}, K., {et~al.} 2014, in Proceedings of
  the SPIE, Vol. 9145, id. 91450

\bibitem[{{Gabuzda} \& {G{\'o}mez}(2001)}]{gabuzda+2001}
{Gabuzda}, D.~C. \& {G{\'o}mez}, J.~L. 2001, \mnras, 320, L49

\bibitem[{{Gurvits} {et~al.}(1992){Gurvits}, {Kardashev}, {Popov}, {Schilizzi},
  {Barthel}, {Pauliny-Toth}, \& {Kellermann}}]{gurvits+1992}
{Gurvits}, L.~I., {Kardashev}, N.~S., {Popov}, M.~V., {et~al.} 1992, \aap, 260,
  82

\bibitem[{{Hirabayashi} {et~al.}(2000){Hirabayashi}, {Hirosawa}, {Kobayashi},
  {Murata}, {Asaki}, {Avruch}, {Edwards}, {Fomalont}, {Ichikawa}, {Kii},
  {Okayasu}, {Wajima}, {Inoue}, {Kawaguchi}, {Chikada}, {Bushimata},
  {Fujisawa}, {Horiuchi}, {Kameno}, {Miyaji}, {Shibata}, {Shen}, {Umemoto},
  {Kasuga}, {Nakajima}, {Takahashi}, {Enome}, {Morimoto}, {Ellis}, {Meier},
  {Murphy}, {Preston}, {Smith}, {Wietfeldt}, {Benson}, {Claussen}, {Flatters},
  {Moellenbrock}, {Romney}, {Ulvestad}, {Langston}, {Minter}, {D'Addario},
  {Dewdney}, {Dougherty}, {Jauncey}, {Lovell}, {Tingay}, {Tzioumis}, {Taylor},
  {Cannon}, {Gurvits}, {Schilizzi}, {Booth}, \& {Popov}}]{hirabayashi+2000}
{Hirabayashi}, H., {Hirosawa}, H., {Kobayashi}, H., {et~al.} 2000, \pasj, 52,
  955

\bibitem[{{Hovatta} {et~al.}(2014){Hovatta}, {Aller}, {Aller}, {Clausen-Brown},
  {Homan}, {Kovalev}, {Lister}, {Pushkarev}, \& {Savolainen}}]{hovatta+2014}
{Hovatta}, T., {Aller}, M.~F., {Aller}, H.~D., {et~al.} 2014, \aj, 147, 143

\bibitem[{{Hovatta} {et~al.}(2012){Hovatta}, {Lister}, {Aller}, {Aller},
  {Homan}, {Kovalev}, {Pushkarev}, \& {Savolainen}}]{hovatta+2012}
{Hovatta}, T., {Lister}, M.~L., {Aller}, M.~F., {et~al.} 2012, \aj, 144, 105

\bibitem[{{Kardashev} {et~al.}(2014{\natexlab{a}}){Kardashev}, {Alakoz},
  {Kovalev}, {Popov}, {Sobolev}, \& {Sokolovsky}}]{kardashev+2014}
{Kardashev}, N.~S., {Alakoz}, A.~V., {Kovalev}, Y.~Y., {et~al.}
  2014{\natexlab{a}}, Proc. of S.A. Lavochkin Association, 3, 4

\bibitem[{{Kardashev} {et~al.}(2013){Kardashev}, {Khartov}, {Abramov},
  {Avdeev}, {Alakoz}, {Aleksandrov}, {Ananthakrishnan}, {Andreyanov},
  {Andrianov}, {Antonov}, {Artyukhov}, {Arkhipov}, {Baan}, {Babakin},
  {Babyshkin}, {Bartel'}, {Belousov}, {Belyaev}, {Berulis}, {Burke},
  {Biryukov}, {Bubnov}, {Burgin}, {Busca}, {Bykadorov}, {Bychkova},
  {Vasil'kov}, {Wellington}, {Vinogradov}, {Wietfeldt}, {Voitsik},
  {Gvamichava}, {Girin}, {Gurvits}, {Dagkesamanskii}, {D'Addario},
  {Giovannini}, {Jauncey}, {Dewdney}, {D'yakov}, {Zharov}, {Zhuravlev},
  {Zaslavskii}, {Zakhvatkin}, {Zinov'ev}, {Ilinen}, {Ipatov}, {Kanevskii},
  {Knorin}, {Casse}, {Kellermann}, {Kovalev}, {Kovalev}, {Kovalenko}, {Kogan},
  {Komaev}, {Konovalenko}, {Kopelyanskii}, {Korneev}, {Kostenko}, {Kotik},
  {Kreisman}, {Kukushkin}, {Kulishenko}, {Cooper}, {Kut'kin}, {Cannon},
  {Larionov}, {Lisakov}, {Litvinenko}, {Likhachev}, {Likhacheva}, {Lobanov},
  {Logvinenko}, {Langston}, {McCracken}, {Medvedev}, {Melekhin}, {Menderov},
  {Murphy}, {Mizyakina}, {Mozgovoi}, {Nikolaev}, {Novikov}, {Novikov},
  {Oreshko}, {Pavlenko}, {Pashchenko}, {Ponomarev}, {Popov}, {Pravin-Kumar},
  {Preston}, {Pyshnov}, {Rakhimov}, {Rozhkov}, {Romney}, {Rocha}, {Rudakov},
  {R{\"a}is{\"a}nen}, {Sazankov}, {Sakharov}, {Semenov}, {Serebrennikov},
  {Schilizzi}, {Skulachev}, {Slysh}, {Smirnov}, {Smith}, {Soglasnov},
  {Sokolovskii}, {Sondaar}, {Stepan'yants}, {Turygin}, {Turygin}, {Tuchin},
  {Urpo}, {Fedorchuk}, {Finkel'shtein}, {Fomalont}, {Fejes}, {Fomina},
  {Khapin}, {Tsarevskii}, {Zensus}, {Chuprikov}, {Shatskaya}, {Shapirovskaya},
  {Sheikhet}, {Shirshakov}, {Schmidt}, {Shnyreva}, {Shpilevskii}, {Ekers}, \&
  {Yakimov}}]{kardashev+2013}
{Kardashev}, N.~S., {Khartov}, V.~V., {Abramov}, V.~V., {et~al.} 2013,
  Astronomy Reports, 57, 153

\bibitem[{{Kardashev} {et~al.}(2014{\natexlab{b}}){Kardashev}, {Kreisman},
  {Pogodin}, {Ponomarev}, {Filippova}, \& {Sheikhet}}]{kardashev+2014b}
{Kardashev}, N.~S., {Kreisman}, B.~B., {Pogodin}, A.~V., {et~al.}
  2014{\natexlab{b}}, Cosmic Research, 52, 332

\bibitem[{{Kemball} {et~al.}(2000){Kemball}, {Flatters}, {Gabuzda},
  {Moellenbrock}, {Edwards}, {Fomalont}, {Hirabayashi}, {Horiuchi}, {Inoue},
  {Kobayashi}, \& {Murata}}]{kemball+2000}
{Kemball}, A., {Flatters}, C., {Gabuzda}, D., {et~al.} 2000, \pasj, 52, 1055

\bibitem[{{Kettenis}(2010)}]{kettenis2010}
{Kettenis}, M. 2010, in 10th European VLBI Network Symposium and EVN Users
  Meeting: VLBI and the New Generation of Radio Arrays, 86

\bibitem[{{Khartov} {et~al.}(2014){Khartov}, {Shirshakov}, {Artyukhov},
  {Kazakevich}, {Vorob'ev}, {Kalashnikov}, {Pogodin}, {Filippova}, \&
  {Komovkin}}]{khartov+2014}
{Khartov}, V.~V., {Shirshakov}, A.~E., {Artyukhov}, M.~I., {et~al.} 2014,
  Cosmic Research, 52, 326

\bibitem[{{Konigl}(1981)}]{koenigl1981}
{Konigl}, A. 1981, \apj, 243, 700

\bibitem[{{Kovalev} {et~al.}(2014){Kovalev}, {Vasil'kov}, {Popov}, {Soglasnov},
  {Voitsik}, {Lisakov}, {Kut'kin}, {Nikolaev}, {Nizhel'skii}, {Zhekanis}, \&
  {Tsybulev}}]{kovalev+2014}
{Kovalev}, Y.~A., {Vasil'kov}, V.~I., {Popov}, M.~V., {et~al.} 2014, Cosmic
  Research, 52, 393

\bibitem[{{Kovalev} {et~al.}(2005){Kovalev}, {Kellermann}, {Lister}, {Homan},
  {Vermeulen}, {Cohen}, {Ros}, {Kadler}, {Lobanov}, {Zensus}, {Kardashev},
  {Gurvits}, {Aller}, \& {Aller}}]{kovalev+2005}
{Kovalev}, Y.~Y., {Kellermann}, K.~I., {Lister}, M.~L., {et~al.} 2005, \aj,
  130, 2473

\bibitem[{{Kravchenko} {et~al.}(2015){Kravchenko}, {Cotton}, \&
  {Kovalev}}]{kravchenko+2015}
{Kravchenko}, E.~V., {Cotton}, W.~D., \& {Kovalev}, Y.~Y. 2015, in IAU
  Symposium, Vol. 313, IAU Symposium, ed. F.~{Massaro}, C.~C. {Cheung},
  E.~{Lopez}, \& A.~{Siemiginowska}, 128--132

\bibitem[{{Lee} {et~al.}(2008){Lee}, {Lobanov}, {Krichbaum}, {Witzel},
  {Zensus}, {Bremer}, {Greve}, \& {Grewing}}]{lee+2008}
{Lee}, S.-S., {Lobanov}, A.~P., {Krichbaum}, T.~P., {et~al.} 2008, \aj, 136,
  159

\bibitem[{{Leppanen} {et~al.}(1995){Leppanen}, {Zensus}, \&
  {Diamond}}]{leppanen+1995}
{Leppanen}, K.~J., {Zensus}, J.~A., \& {Diamond}, P.~J. 1995, \aj, 110, 2479

\bibitem[{{Levy} {et~al.}(1986){Levy}, {Linfield}, {Ulvestad}, {Edwards},
  {Jordan}, {di Nardo}, {Christensen}, {Preston}, {Skjerve}, \&
  {Blaney}}]{levy+1986}
{Levy}, G.~S., {Linfield}, R.~P., {Ulvestad}, J.~S., {et~al.} 1986, Science,
  234, 187

\bibitem[{{Lisakov} {et~al.}(2014){Lisakov}, {Voinakov}, {Syrov}, {Sokolov},
  {Dobrynin}, {Shatsky}, {Kamaldinova}, {Sosnovtsev}, {Ryabogin},
  {Vyunitskaya}, \& {Filippova}}]{lisakov+2014}
{Lisakov}, M.~M., {Voinakov}, S.~M., {Syrov}, A.~S., {et~al.} 2014, Cosmic
  Research, 52, 365

\bibitem[{{Lister} {et~al.}(2013){Lister}, {Aller}, {Aller}, {Homan},
  {Kellermann}, {Kovalev}, {Pushkarev}, {Richards}, {Ros}, \&
  {Savolainen}}]{lister+2013}
{Lister}, M.~L., {Aller}, M.~F., {Aller}, H.~D., {et~al.} 2013, \aj, 146, 120

\bibitem[{{Lobanov}(1998)}]{lobanov1998}
{Lobanov}, A.~P. 1998, \aap, 330, 79

\bibitem[{{Lobanov}(2015)}]{lobanov2015}
{Lobanov}, A.~P. 2015, \aap, 574, A84

\bibitem[{{Lobanov} {et~al.}(2003){Lobanov}, {Hardee}, \&
  {Eilek}}]{lobanov+2003}
{Lobanov}, A.~P., {Hardee}, P., \& {Eilek}, J. 2003, \nar, 47, 629

\bibitem[{{Lobanov} {et~al.}(2000){Lobanov}, {Krichbaum}, {Graham}, {Witzel},
  {Kraus}, {Zensus}, {Britzen}, {Greve}, \& {Grewing}}]{lobanov+2000}
{Lobanov}, A.~P., {Krichbaum}, T.~P., {Graham}, D.~A., {et~al.} 2000, \aap,
  364, 391

\bibitem[{{Murphy}(1991)}]{murphy1991}
{Murphy}, D.~W. 1991, in Astronomical Society of the Pacific Conference Series,
  Vol.~19, IAU Colloq. 131: Radio Interferometry. Theory, Techniques, and
  Applications, ed. T.~J. {Cornwell} \& R.~A. {Perley}, 107

\bibitem[{{Murphy} {et~al.}(1994){Murphy}, {Yakimov}, {Kobayashi}, {Taylor}, \&
  {Fejes}}]{murphy+1994}
{Murphy}, D.~W., {Yakimov}, V., {Kobayashi}, H., {Taylor}, A.~R., \& {Fejes},
  I. 1994, in VLBI TecnologyY: Progress and Future Observational Possibilities,
  ed. T.~{Sasao}, S.~{Manabe}, O.~{Kameya}, \& M.~{Inoue}, 34--38

\bibitem[{{Oppermann} {et~al.}(2012){Oppermann}, {Junklewitz}, {Robbers},
  {Bell}, {En{\ss}lin}, {Bonafede}, {Braun}, {Brown}, {Clarke}, {Feain},
  {Gaensler}, {Hammond}, {Harvey-Smith}, {Heald}, {Johnston-Hollitt}, {Klein},
  {Kronberg}, {Mao}, {McClure-Griffiths}, {O'Sullivan}, {Pratley}, {Robishaw},
  {Roy}, {Schnitzeler}, {Sotomayor-Beltran}, {Stevens}, {Stil}, {Sunstrum},
  {Tanna}, {Taylor}, \& {Van Eck}}]{oppermann+2012}
{Oppermann}, N., {Junklewitz}, H., {Robbers}, G., {et~al.} 2012, \aap, 542, A93

\bibitem[{{Osmer} {et~al.}(1994){Osmer}, {Porter}, \& {Green}}]{osmer+1994}
{Osmer}, P.~S., {Porter}, A.~C., \& {Green}, R.~F. 1994, \apj, 436, 678

\bibitem[{{O'Sullivan} {et~al.}(2011){O'Sullivan}, {Gabuzda}, \&
  {Gurvits}}]{osullivan+2011}
{O'Sullivan}, S.~P., {Gabuzda}, D.~C., \& {Gurvits}, L.~I. 2011, \mnras, 415,
  3049

\bibitem[{{Owen} {et~al.}(1989){Owen}, {Hardee}, \& {Cornwell}}]{owen+1989}
{Owen}, F.~N., {Hardee}, P.~E., \& {Cornwell}, T.~J. 1989, \apj, 340, 698

\bibitem[{{Pashchenko} {et~al.}(2015){Pashchenko}, {Kovalev}, \&
  {Voitsik}}]{pashchenko+2015}
{Pashchenko}, I.~N., {Kovalev}, Y.~Y., \& {Voitsik}, P.~A. 2015, Cosmic
  Research, 53, 199

\bibitem[{{Planck Collaboration} {et~al.}(2015){Planck Collaboration}, {Ade},
  {Aghanim}, {Arnaud}, {Ashdown}, {Aumont}, {Baccigalupi}, {Banday},
  {Barreiro}, {Bartlett}, \& et~al.}]{planck2015}
{Planck Collaboration}, {Ade}, P.~A.~R., {Aghanim}, N., {et~al.} 2015, ArXiv
  e-prints, 1502.01589

\bibitem[{{Pushkarev} {et~al.}(2012){Pushkarev}, {Hovatta}, {Kovalev},
  {Lister}, {Lobanov}, {Savolainen}, \& {Zensus}}]{pushkarev+2012}
{Pushkarev}, A.~B., {Hovatta}, T., {Kovalev}, Y.~Y., {et~al.} 2012, \aap, 545,
  A113

\bibitem[{{Pushkarev} {et~al.}(2009){Pushkarev}, {Kovalev}, {Lister}, \&
  {Savolainen}}]{pushkarev+2009}
{Pushkarev}, A.~B., {Kovalev}, Y.~Y., {Lister}, M.~L., \& {Savolainen}, T.
  2009, \aap, 507, L33

\bibitem[{{Shepherd}(2011)}]{shepherd2011}
{Shepherd}, M. 2011, {Difmap: Synthesis Imaging of Visibility Data},
  Astrophysics Source Code Library

\bibitem[{{Shepherd}(1997)}]{shepherd1997}
{Shepherd}, M.~C. 1997, in Astronomical Society of the Pacific Conference
  Series, Vol. 125, Astronomical Data Analysis Software and Systems VI, ed.
  G.~{Hunt} \& H.~{Payne}, 77

\bibitem[{{Sparks} {et~al.}(1996){Sparks}, {Biretta}, \&
  {Macchetto}}]{sparks+1996}
{Sparks}, W.~B., {Biretta}, J.~A., \& {Macchetto}, F. 1996, \apj, 473, 254

\bibitem[{{Taylor} {et~al.}(2009){Taylor}, {Stil}, \& {Sunstrum}}]{taylor+2009}
{Taylor}, A.~R., {Stil}, J.~M., \& {Sunstrum}, C. 2009, \apj, 702, 1230

\bibitem[{{Torrealba} {et~al.}(2012){Torrealba}, {Chavushyan},
  {Cruz-Gonz{\'a}lez}, {Arshakian}, {Bertone}, \&
  {Rosa-Gonz{\'a}lez}}]{torrealba+2012}
{Torrealba}, J., {Chavushyan}, V., {Cruz-Gonz{\'a}lez}, I., {et~al.} 2012,
  \rmxaa, 48, 9

\bibitem[{{Turygin}(2014)}]{turygin2014}
{Turygin}, M.~S. 2014, Cosmic Research, 52, 403

\bibitem[{{Xu} {et~al.}(1995){Xu}, {Readhead}, {Pearson}, {Polatidis}, \&
  {Wilkinson}}]{xu+1995}
{Xu}, W., {Readhead}, A.~C.~S., {Pearson}, T.~J., {Polatidis}, A.~G., \&
  {Wilkinson}, P.~N. 1995, \apjs, 99, 297

\bibitem[{{Zakhvatkin} {et~al.}(2014){Zakhvatkin}, {Ponomarev}, {Stepan'yants},
  {Tuchin}, \& {Zaslavskiy}}]{zakhvatkin+2014}
{Zakhvatkin}, M.~V., {Ponomarev}, Y.~N., {Stepan'yants}, V.~A., {Tuchin},
  A.~G., \& {Zaslavskiy}, G.~S. 2014, Cosmic Research, 52, 342

\bibitem[{{Zhuravlev}(2014)}]{zhuravlev2014}
{Zhuravlev}, V.~I. 2014, ArXiv e-prints, 1404.2430

\end{thebibliography}

\end{document}